\documentclass[showpacs,preprintnumbers,eqsecnum,nofootinbib,twocolumn,prd]{revtex4}  

\usepackage{graphicx}  
\usepackage{subfigure}
\usepackage{latexsym}
\usepackage{amssymb}

\def\beq{\begin{equation}}
\def\eeq{\end{equation}}
\def\rmd{{\rm d}}

\begin{document}

\title{Tidal invariants along the world line of an extended body in the Kerr spacetime}

\author{Donato Bini and Andrea Geralico}
  \affiliation{
Istituto per le Applicazioni del Calcolo ``M. Picone,'' CNR, I-00185 Rome, Italy
}

\date{\today}

\begin{abstract}
An extended body orbiting a compact object undergoes tidal deformations by the background gravitational field. 
Tidal invariants built up with the Riemann tensor and their derivatives evaluated along the world line of the body are essential tools to investigate both geometrical and physical properties of the tidal interaction.
For example, one can determine the tidal potential in the neighborhood of the body by constructing a body-fixed frame, which requires Fermi-type coordinates attached to the body itself, the latter being in turn related to the spacetime metric and curvature along the considered world line.
Similarly, in an effective field theory description of extended bodies finite size effects are taken into account by adding to the point mass action certain non-minimal couplings which involve integrals of tidal invariants along the orbit of the body.
In both cases such a computation of tidal tensors is required.
Here we consider the case of a spinning body also endowed with a non-vanishing quadrupole moment in a Kerr spacetime.
The structure of the body is modeled by a multipolar expansion around the \lq\lq center of mass line'' according to the Mathisson-Papapetrou-Dixon model truncated at the quadrupolar order. The quadrupole tensor is assumed to be quadratic in spin, accounting for rotational deformations. The behavior of tidal invariants of both electric and magnetic type is discussed in terms of gauge-invariant quantities when the body is moving along a circular orbit as well as in the case of an arbitrary (equatorial) motion.
The analysis is completed by examining the associated eigenvalues and eigenvectors of the tidal tensors.
The limiting situation of the Schwarzschild solution is also explored both in the strong field regime and in the weak field limit. 
\end{abstract}


\maketitle

\section{Introduction}

Tidal interactions are expected to play a significant role in driving the dynamics of binary systems. 
For instance, gravitational waves emitted in the late stages of coalescing neutron star binaries should contain an imprint of the two-body tidal interaction. Several techniques have been developed so far to study relativistic tidal problems in the strong field regime at different levels of approximation.
In the last decade we have seen the fruitful interaction of theoretical approaches, like the post-Newtonian (PN) and post-Minkowskian approximations, the gravitational self-force (SF) and the effective one-body (EOB) formalisms, besides the numerical relativity simulations \cite{Flanagan:2007ix,Read:2009yp,Baiotti:2010xh,Bernuzzi:2012ci,Damour:2012yf,Bernuzzi:2011aq,Read:2013zra,DelPozzo:2013ala,Hotokezaka:2013mm,Radice:2013hxh,Bernuzzi:2014kca,Bernuzzi:2014owa}.
Unfortunately, analyzing the strong field aspects of the dynamics of compact binaries is not an easy task, either from a theoretical point of view or from a fully numerical one.
The state-of-the-art of our knowledge of tidal interactions in non-spinning comparable mass binary systems is substantially limited to the second PN approximation \cite{Bini:2012gu}, while in the limiting situation of an extreme-mass-ratio binary systems Dolan et al. \cite{Dolan:2014pja} have obtained last year high-precision numerical results in the framework of linear perturbation theory, immediately followed by high-order PN accuracy theoretical computations by Bini and Damour \cite{Bini:2014zxa,Bini:2015bla,Bini:2015mza} (up to the 7.5PN level first, and later to 10.5PN), showing helpful synergies between analytical and numerical approaches to this subject.
Noticeably, very recently Kavanagh, Ottewill and Wardell \cite{Kavanagh:2015lva} as well as Johnson-McDaniel, Shah and Whiting \cite{Johnson-McDaniel:2015vva} have been able to raise the PN precision of these computations to the level of 21.5PN.
Finite size effects induced by the shape of the bodies modify the dynamics of compact binary systems starting at the 5PN level, and have a counterpart in the notion of Love numbers (measuring the tidal polarizability of the bodies) \cite{Damour:1982wm,hin1,Bin-poi,dam-nag10}.
The case of tidal interactions among spinning bodies has not received yet enough attention in the literature, and is the object of the present study.

When the mass of one body is much smaller than that of its companion, its dynamics can be treated as the motion of an extended body in a given gravitational field due to the body of higher mass. This is the case of a star orbiting a compact object, which gets deformed by the tidal field associated with their mutual interaction \cite{fishbone,hiscock,marck,carter,marck2,shibata,frolov,mino}. 
Tidal effects can be studied by constructing a body-fixed frame adapted to the timelike geodesic path along with the (center of mass of the) star is assumed to move under the action of a tidal potential, in terms of Fermi-type coordinates. 
These coordinates are related by definition to the background spacetime metric and curvature tensor evaluated along the world line of the body itself. 
Another approach consists in developing an effective field theory description of the extended body. In this case finite size effects are taken into account by adding to the point mass action certain non-minimal couplings which involve integrals of tidal invariants performed (again) along the path of the body. We stress that in both cases the computation of tidal tensors is required only along the body's world line, whose knowledge should be supplied as an additional information.
This general framework still holds in perturbation theory, where the backreaction of the body on the background geometry is taken into account, as well as in the PN approximation. 
We will adopt such an effective action description of tidal effects to investigate the properties of tidal invariants associated with the world line of an extended body endowed with both dipolar and quadrupolar structure moving in the equatorial plane of a Kerr spacetime.
We will focus on quadratic and cubic invariants of both electric and magnetic type, leaving for a future study more involved ones, like differential invariants constructed from the covariant derivative of the curvature tensor.

Let us make more precise some notational aspect.
The masses of the two gravitationally interacting bodies are denoted by $m_1$ and $m_2$, with the convention that $m_1\le m_2$.
The body of mass $m_1$ is endowed with spin $S_1$ and that of mass $m_2$ is endowed with spin $S_2$.
We define, in a standard way, the total mass of the system ($\mathcal M$), the reduced mass ($\mu$) and the symmetric mass-ratio ($\nu$) as
\begin{eqnarray}
\label{eq:1.1}
\mathcal M &\equiv& m_1+m_2\,,\qquad
\mu\equiv\frac{m_1 m_2}{\mathcal M}\,,\nonumber\\
\nu &\equiv& \frac{\mu}{\mathcal M}=\frac{m_1m_2}{(m_1+m_2)^2}
\,.
\end{eqnarray}
Other dimensionless mass ratios are often used, e.g.,
\beq
\label{eq:1.2}
q\equiv \frac{m_1}{m_2}\le 1\,,\quad 
X_1\equiv \frac{m_1}{\mathcal M}\le \frac12\,,\quad
X_2\equiv \frac{m_2}{\mathcal M}= 1-X_1\,,
\eeq
with the links
\beq
\label{eq:1.3}
X_1=\frac{q}{1+q}\,,\quad
X_2=\frac{1}{1+q}\,,\quad
\nu=X_1 X_2=\frac{q}{(1+q)^2}
\,.
\eeq
In the limit of small mass-ratio ($q\ll 1$), we have $\nu \simeq X_1\simeq q$ and $X_2 \simeq 1-q$.
Finally, one usually defines $\chi_{1,2}\equiv S_{1,2}/m_{1,2}^2$ as the dimensionless spin variable associated with each body.
However, alternative definitions for the dimensionless spin are also used in the literature, e.g., $S_{1,2}/(m_1m_2)$ and $S_{1,2}/{\mathcal M}^2$.

\section{Tidal invariants}

In an effective field theory description of extended objects, finite size effects are treated by increasing the (leading-order) point-mass action \cite{Damour:1982wm}
\beq
\label{eq:2.1}
S_0=\int \frac{\rmd^4 x}{c}\, \frac{c^4}{16 \pi G}\sqrt{-g} R-\sum_A \int m_A c^2 \rmd\tau_A\,,
\eeq
by additional, non-minimal, couplings involving higher-order derivatives of the field evaluated along the world line of the body \cite{Damour:1995kt,Damour:1998jk,Goldberger:2004jt,porto06,porto06prl,porto08,Levi:2014gsa,Levi:2015msa}. 
Here $\rmd\tau_A=-(U_A)_\mu \rmd y_A^\mu$ is the (dimensionally regularized) proper time along the world line $y_A^\mu(\tau_A)$ of body $A$, with $4$-velocity $U_A^\mu = \rmd y_A^\mu/\rmd\tau_A$. The body $A$ feels the gravitational field of the whole interacting $N$-body system, which can be described by a suitably defined \lq\lq external metric'' \cite{Zhang86,Damour:1990pi,Damour:1991yw}. 
Non-minimal couplings are expressed in terms of two types of tidal tensors computed in this metric: the gravitoelectric $G_L^A(\tau_A)\equiv G^A_{a_1\ldots a_l}(\tau_A)$, and gravitomagnetic $H_L^A(\tau_A)\equiv H^A_{a_1\ldots a_l}(\tau_A)$, symmetric trace-free (spatial) tensors, together with their proper time derivatives. (The spatial indices $a_i=1,2,3$ refer to body-fixed coordinates $X^\alpha_A=(c\tau_A, X_A^a)$, attached to body $A$.) In terms of these tidal tensors, the most general world line action has the form \cite{Damour:1990pi,Damour:1991yw,Damour:1992qi,Damour:1993zn}
\beq
S_{\rm non-min} 
=\sum_{A}S_{\rm non-min} ^{A}\,,
\eeq
with
\begin{eqnarray}
\label{eq:2.2}
S_{\rm non-min} ^{A}&=&  
\frac{1}{4} \, \mu_A^{(2)} \int \rmd\tau_A \, G_{\alpha\beta}^A \, G_A^{\alpha\beta}\nonumber\\
&&
+\frac{1}{6 \, c^2} \, \sigma_A^{(2)} \int \rmd\tau_A \, H_{\alpha\beta}^A \, H_A^{\alpha\beta} \nonumber \\
&& 
+\frac{1}{12} \, \mu_A^{(3)} \int \rmd\tau_A \, G_{\alpha\beta\gamma}^A \, G_A^{\alpha\beta\gamma} \nonumber \\
&& 
+\frac{1}{4 \, c^2} \, \mu'^{(2)}_A \int \rmd\tau_A \dot G_{\alpha\beta}^A\dot G_A^{\alpha\beta}  +\ldots  \, ,
\end{eqnarray}
where $G^A_{\alpha\beta\gamma}={\rm Sym}_{\alpha\beta\gamma}\,(P(U_A)_\alpha^\mu \nabla_\mu R_{\beta \rho \gamma \nu})U_A^\rho U_A^\nu $ [with $P(U_A)=g+U_A\otimes U_A$ projecting orthogonally to $U_A$,] $\dot G_A^{\alpha\beta}\equiv U_A^{\mu} \nabla_{\mu} G_{\alpha\beta}^A$, and $\mu_A^{(2)}$, $\sigma_A^{(2)}$, $\mu_A^{(3)}$, $\mu'^{(2)}_A$ are tidal coefficients. 
Higher-order invariants involve higher-than-quadratic tidal scalars, e.g.,
cubic in $G_{ab}^A$ 
\beq
\label{eq:2.3}
\int \rmd\tau_A G^A{}_{ab}G^A{}^{bc}G^A{}_{c}{}^{a}\,.
\eeq
As it follows from the above expressions, all these tidal tensors should be then evaluated along the world line of the body itself.
This makes the problem more involved than the computation of coordinate invariants (simply built up through the spacetime metric), since the explicit solution for the world line of the body is also needed.

We shall focus here on the simplest invariants associated with the quadrupolar electric-type and magnetic-type tidal tensors $G_{ab}$, $H_{ab}$. The latter are related as follows to the spatial components of the \lq\lq electric" and \lq\lq magnetic" parts of the Riemann tensor (evaluated, with dimensional regularization, along the considered world line)
\beq
\label{eq:2.4}
G^A_{\alpha\beta}\equiv -{\mathcal E}^A_{\alpha\beta}(U_A)\,,\qquad
H^A_{\alpha\beta} \equiv  2\, c\, {\mathcal B}^A_{\alpha\beta}(U_A)\,,
\eeq
where ${\mathcal E}^A_{\alpha\beta}(U_A)$ and ${\mathcal B}^A_{\alpha\beta}(U_A)$ are defined as
\begin{eqnarray}
\label{riemann_em}
{\mathcal E}^A_{\alpha\beta}(U_A)&=& R_{\alpha\mu\beta\nu}U_A^\mu U_A^\nu\,,\nonumber\\
{\mathcal B}^A_{\alpha\beta}(U_A)&=&[R^*]_{\alpha\mu\beta\nu}U_A^\mu U_A^\nu\,,
\end{eqnarray}
the symbol $^*$ denoting the spacetime dual of a tensor, as standard.
Here, we are interested in the tidal invariants of the body labeled 1 (with mass $m_1$), member of a binary system (i.e., $N=2$).
For ease of notation, we shall henceforth often suppress the body label $A=1$.
Furthermore, we shall set $G=c=1$. 
The associated non-minimal world line action (\ref{eq:2.2}) thus writes as
\begin{eqnarray}
\label{eq:2.2n}
S_{1\rm non-min}
&=&  \frac{1}{4} \, \mu_1^{(2)} \int \rmd\tau_1 \, {\mathcal T}_E(U_1)\nonumber\\
&&
+\frac{2}{3} \, \sigma_1^{(2)} \int \rmd\tau_1 {\mathcal T}_B(U_1) \nonumber\\
&&
+\frac{1}{12} \, \mu_1^{(3)} \int \rmd\tau_1 \, {\mathcal T}_G(U_1)\nonumber\\
&&
+\frac{1}{4} \, \mu'^{(2)}_1 \int \rmd\tau_1 {\mathcal T}_{\dot G}(U_1)+\ldots\,,  
\end{eqnarray}
where ${\mathcal T}_E(U_1)\equiv {\rm Tr}\,[{\mathcal E}(U_1)]^2$, ${\mathcal T}_B(U_1)\equiv {\rm Tr}\,[{\mathcal B}(U_1)]^2$, ${\mathcal T}_G(U_1)\equiv G_{\alpha\beta\gamma}^1 \,G_1^{\alpha\beta\gamma}$, ${\mathcal T}_{\dot G}(U_1)\equiv \dot G_{\alpha\beta}^1\dot G_1^{\alpha\beta}$.\footnote{
The quadratic tidal invariants of electric and magnetic type are denoted in Ref. \cite{Bini:2014zxa} by $J_{1e^2}$ and $J_{1b^2}$, respectively.
}

The quadrupolar electric-type tidal tensor (\ref{eq:2.4}), in non-spinning comparable mass binary systems,  has been computed to 1PN fractional accuracy in Refs. \cite{Damour:1992qi,Damour:1993zn} (see also Refs. \cite{Vines:2010ca,Taylor:2008xy} fore more details). Ref. \cite{JohnsonMcDaniel:2009dq} has also computed to 1PN accuracy the octupolar electric-type tidal tensor, $G_{abc}$, and the quadrupolar magnetic-type tidal tensor $H_{ab}\sim {\mathcal B}_{ab}$. The significantly more involved calculation of tidal effects, along general orbits, but still in the case of non-spinning binary systems at the 2PN fractional accuracy has been done in Ref. \cite{Bini:2012gu}. 

Let us quote the values of the 2PN-accurate tidal invariants computed in Ref. \cite{Bini:2012gu} for the simplest case of two bodies moving along spatially circular orbits with (coordinate time) constant angular velocity $\Omega$,\footnote{
The coordinate time angular velocity will be denoted by $\zeta$ in the next section, leaving the notation $\Omega$ for the proper time angular velocity. 
} expressed in a gauge-invariant way in terms of the symmetric, dimensionless frequency parameter 
$x=[(m_1+m_2)\Omega]^{2/3}$.
They can be cast in the following form \cite{Bini:2012gu}
\begin{eqnarray}
\label{eq:2.8}
{\mathcal M}^4{\mathcal T}_E(U_1)
&=& 6X_2^2x^6 \left[\frac{1-3x+3x^2}{(1-3x)^2} + (2X_1^2-X_1)x\right.\nonumber\\  
&&\left. 
+\left(\frac53 X_1^4-X_1^3+\frac{787}{84}X_1^2+\frac14 X_1\right)x^2\right.\nonumber\\  
&&\left. 
+O_{X_1}(x^3)\right]\,,\nonumber\\
{\mathcal M}^4{\mathcal T}_B(U_1)
&=& 18X_2^2x^7 \left[\frac{1-2x}{(1-3x)^2} +\left(\frac{10}{3} X_1^2-2 X_1 \right)x\right.\nonumber\\  
&&\left. 
+O_{X_1}(x^2)\right]\,, 
\end{eqnarray}
where the notation $O_a(x^n)$ denotes a term which vanishes with $a$ and which is $O(x^n)$.
In addition, let us recall  that $x$ is related to the body-dissymmetric (but SF motivated) frequency parameter $y=(m_2\Omega)^{2/3}$ by $x=(1+q)^{2/3}y$.

We will evaluate below the tidal invariants ${\mathcal T}_E(U_1)$, ${\mathcal T}_B(U_1)$ and ${\rm Tr}\,[{\mathcal E}(U_1)]^3$ along the world line ${\mathcal L}_1$ of the smaller mass $m_1$ with spin $S_1$, which we will assume to be tidally deformed by its own spin.
We will not consider here more involved tidal invariants, like differential invariants constructed from the covariant derivative of the curvature tensor.
The latter play a role, e.g., in the context of the tidal interaction between ordinary stars and compact objects orbiting a black hole analyzed by using Fermi coordinate approximated tidal potentials, as discussed in Ref. \cite{mino}. The basic assumption there is that the presence of the star does not perturb the background field, and it can be described as a self-gravitating Newtonian fluid, whose center of mass moves along a timelike geodesic path. The tidal field due to the black hole is then computed from the Riemann tensor in terms of the geodesic deviation equation.  

The tidal invariants defined above all have an intrinsic observer-dependent meaning (see Eqs. (\ref{eq:2.4}) and (\ref{riemann_em})).
In previous papers we have investigated the role of the observer measuring quadratic tidal invariants both in the Kerr spacetime \cite{Bini:2012zzb} and in the spacetime of a rotating deformed mass \cite{Bini:2012zze}.
We have considered there different families of observers which have a special geometrical and physical meaning as well as observers carrying an intrinsic spin.
In particular, we have explored the family of stationary circularly rotating observers in the equatorial plane, including ZAMOs and geodesic observers, showing that no observer within this family can measure a vanishing electric tidal indicator, whereas it is the family of Carter's observers that measures a zero magnetic one.

When the internal structure of the moving body is taken into account, the tidal invariants will contain information about it. We will consider below an extended body endowed with both dipolar and quadrupolar structure as described by the Mathisson-Papapetrou-Dixon (MPD) model \cite{math37,papa51,tulc59,dixon64,dixon69,dixon70,dixon73,dixon74}. The quadrupole tensor will be assumed to be spin-induced, i.e., proportional to the trace-free part of the square of the spin tensor by a constant parameter which is characteristic of the body under consideration. For instance, for neutron stars it depends on the equation of state \cite{poisson}. 
We will consider the special case of circular motion as well as quasi-circular orbits, i.e., orbits which deviate from the reference circular geodesic motion due to both the spin-curvature force and the quadrupolar force.

\section{Equatorial motion in a Kerr spacetime}
 
In standard Boyer-Lindquist coordinates the Kerr metric writes as
\begin{eqnarray}
\rmd s^2 &=& -\left(1-\frac{2Mr}{\Sigma}\right)\rmd t^2 -\frac{4aMr}{\Sigma}\sin^2\theta\rmd t\rmd\phi\nonumber\\  
&&
+\frac{\Sigma}{\Delta}\rmd r^2+\Sigma\rmd \theta^2+\frac{\Lambda}{\Sigma}\sin^2 \theta \rmd \phi^2\,,
\end{eqnarray}
where $\Delta=r^2-2Mr+a^2$, $\Sigma=r^2+a^2\cos^2\theta$ and $\Lambda = (r^2+a^2)^2-\Delta a^2\sin^2 \theta$. Here $M$ and $a\le M$, as standard too, are the total mass and the specific angular momentum characterizing the spacetime.
The event horizons are located at $r_\pm=M\pm\sqrt{M^2-a^2}$. 

Introduce the zero angular momentum observer (ZAMO) family of fiducial observers with 4-velocity $n$ orthogonal to the time coordinate hypersurfaces
\beq
\label{n}
n=N^{-1}(\partial_t-N^{\phi}\partial_\phi)\,,
\eeq
where $N=(-g^{tt})^{-1/2}=\left[\Delta\Sigma/\Lambda\right]^{1/2}$ and $N^{\phi}=g_{t\phi}/g_{\phi\phi}=-2aMr/\Lambda$ are the lapse function and only nonvanishing component of the shift vector field respectively. 
The ZAMOs are accelerated and locally nonrotating in the sense that their vorticity vector vanishes; they have also a nonzero expansion tensor.
A suitable orthonormal frame adapted to the ZAMOs is given by
\begin{eqnarray}
\label{zamoframe}
e_{\hat t}&=&n\,, \qquad
e_{\hat r}=\frac1{\sqrt{g_{rr}}}\partial_r\,, \nonumber\\
e_{\hat \theta}&=&\frac1{\sqrt{g_{\theta \theta }}}\partial_\theta\,, \qquad
e_{\hat \phi}=\frac1{\sqrt{g_{\phi \phi }}}\partial_\phi\,.
\end{eqnarray}

The electric and magnetic quadratic tidal invariants introduced in the previous section are simply related by \cite{Bini:2012zzb} 
\beq
\label{relTEH}
{\mathcal T}_E(U)-{\mathcal T}_B(U)=\frac{6M^2}{r^6}\,,
\eeq
where $U$ denotes the unit tangent vector to a given world line ${\mathcal L}$.
For instance, for ZAMOs ($U=n$) we find
\begin{eqnarray}
\label{tidalindzamo}
{\mathcal T}_E(n)&=&\frac{6M^2}{r^6}+{\mathcal T}_B(n)\,,\nonumber\\
{\mathcal T}_B(n)&=&\frac{18a^2M^2}{r^8}\frac{(r^2+a^2)^2\Delta}{(2a^2M+r^3+a^2r)^2}\,.
\end{eqnarray}
This invariance property has been proven in Ref. \cite{Bini:2012zzb} to hold for any given family of equatorial circularly rotating observers. It has a simple explanation in terms of the Kretschmann invariant $K$ of the spacetime, since the difference between the tidal invariants is just proportional to $K$ evaluated at $\theta=\pi/2$. Actually, it is possible to show that the validity of Eq. (\ref{relTEH}) can be extended to an arbitrarily moving observer, not only in the equatorial plane.
In fact, let $\{E_{\hat 0}\equiv U,E_{\hat i}\}$ be an arbitrary orthonormal frame adapted to a generic observer congruence with 4-velocity field $U$ parametrized by the proper time $\tau$. 
From the definition of the Kretschmann invariant
\beq
\label{Kdef}
K=R_{\alpha\beta\gamma\delta}R^{\alpha\beta\gamma\delta}\,,
\eeq
passing to frame components leads (in vacuum) to 
\begin{eqnarray}
K&=&4R_{\hat 0\hat i\hat 0\hat j}R^{\hat 0\hat i\hat 0\hat j}-4R_{\hat 0\hat k\hat i\hat j}R_{\hat 0}{}^{\hat k\hat i\hat j}+R_{\hat i\hat j\hat k\hat l}R^{\hat i\hat j\hat k\hat l}\nonumber\\
&=&8[{\mathcal T}_E(U)-{\mathcal T}_B(U)]\,,
\end{eqnarray}
where $R_{\hat 0\hat i\hat 0\hat j}={\mathcal E}(U)_{\hat i\hat j}$ and $R_{\hat k\hat 0}{}^{\hat i\hat j}={\mathcal B}(U)_{\hat k\hat r}\epsilon^{\hat r\hat i\hat j}$, $\epsilon_{\hat i\hat j\hat k}$ being the Levi-Civita alternating symbol associated with the spatial orthonormal frame $\{E_{\hat i}\}$. 
Therefore, the relation $K=8[{\mathcal T}_E(U)-{\mathcal T}_B(U)]$ holds for any observer's world line, provided that the Kretschmann invariant is evaluated along the chosen world line too.
For instance, for an arbitrarily moving observer in the equatorial plane the previous relation reduce to Eq. (\ref{relTEH}) with $r=r(\tau)$ as given by the parametric equations of the orbit (e.g., $r=r_0=$ constant in the case of circular orbits).

For a later convenience, we will use a tilde notation for mass-rescaled dimensionless quantities, e.g.,
$\tilde {\mathcal T}_{E,B}(U)=M^4 {\mathcal T}_{E,B}(U)$.

\subsection{Circular orbits: a brief overview of their geometrical characterization}
\label{circ}

Consider a family of uniformly rotating timelike circular orbits at a given fixed radius on the equatorial plane with  4-velocity vector $U_{\rm (circ)}$. It can be parametrized equivalently either by the constant angular velocity $\zeta$ with respect to infinity or by the constant relative velocity $\nu$ with respect to the ZAMOs (defining the usual Lorentz factor $\gamma=(1-\nu^2)^{-1/2}$) 
as follows 
\beq
\label{orbita}
U_{\rm (circ)}=\Gamma [\partial_t +\zeta \partial_\phi ]
 =\gamma [n +\nu e_{\hat \phi}]
\,,
\eeq
where $\Gamma$ is a normalization factor such that $U_{\rm (circ)\,\alpha}U_{\rm (circ)}^\alpha =-1$ and hence
\beq
\Gamma =\left[ N^2-g_{\phi\phi}(\zeta+N^{\phi})^2 \right]^{-1/2}
       =\frac{\gamma}{N}\,,
\eeq
with
\beq
\label{zetavsnu}
\zeta=-N^{\phi}+\frac{N}{\sqrt{g_{\phi\phi}}}\, \nu\ \,,\qquad
\nu=\frac{\sqrt{g_{\phi\phi}}}{N} (\zeta+N^{\phi}) 
\,.
\eeq
The parametric equations of the orbit are then given by
\beq
\label{casocircorb}
t=t_0+\Gamma \tau\,,\quad 
r=r_0\,,\quad 
\theta=\frac{\pi}{2}\,,\quad
\phi=\phi_0+\Omega \tau\,,
\eeq
with proper time angular velocity $\Omega$ and coordinate time angular velocity $\zeta$ related by $\Omega=\Gamma\zeta$.
For a later use, we introduce the dimensionless coordinate time angular velocity
\beq
\label{ydef}
y\equiv (M\zeta)^{2/3}\,,
\eeq
and a spacelike unit vector $\bar U_{\rm (circ)}$ within the Killing 2-plane which is orthogonal to $U_{\rm (circ)}$ given by
\beq
\label{barU}
\bar U_{\rm (circ)}=\bar \Gamma [\partial_t +\bar \zeta \partial_\phi ]
={\rm sgn}(\nu)\gamma [\nu e_{\hat t} +e_{\hat \phi}]\,,
\eeq
with
\beq
\bar \zeta=-N^{\phi}+\frac{N}{\sqrt{g_{\phi\phi}}}\, \frac{1}{\nu}\,, \qquad 
\bar \Gamma=\Gamma|\nu|\,.
\eeq

Co-rotating $(+)$ and counter-rotating $(-)$ geodesics $U_\pm$ (with respect to the rotation of the background source,
which is clockwise assuming $a>0$) are characterized by the following angular and linear velocities
\beq
\zeta_{\pm}
=\left[a\pm (M/r^3)^{-1/2}\right]^{-1}\,, \quad
\nu_\pm 
=\frac{a^2\mp2a\sqrt{Mr}+r^2}{\sqrt{\Delta}(a\pm r\sqrt{r/M})}\,,
\eeq 
respectively, and the associated normalization factor $\Gamma(\zeta_\pm)\equiv \Gamma_\pm$ given by
\beq
\Gamma_\pm=\frac{r^{3/2}\pm a\sqrt{M}}{r^{3/4}\left[\sqrt{r}(r-3M)\pm 2a\sqrt{M}\right]^{1/2}}\,.
\eeq
Furthermore, the orthogonal unit vector $\bar U_{\pm}$ is given by
\beq
\label{barUgeo}
\bar U_{\pm}=\bar \Gamma_\pm [\partial_t +\bar \zeta_{\pm} \partial_\phi ]=\pm\gamma_{\pm} [\nu_{\pm} e_{\hat t} + e_{\hat \phi}]\,,
\eeq
where the $\pm$ signs are correlated with those in $U_{\pm}$, $\bar\Gamma_\pm=\Gamma_{\pm}|\nu_{\pm}|$ and 
\beq
\bar\zeta_\pm=\pm\frac{r^{3/2}-2M\sqrt{r}\pm a\sqrt{M}}{\sqrt{M}(r^2\mp2a\sqrt{Mr}+a^2)}\,.
\eeq
In the static case $\zeta_\pm\to\pm\zeta_K$, $\nu_\pm\to\pm\nu_K$, $\Gamma_\pm\to\Gamma_K$ and $\bar\zeta_\pm\to\pm\zeta_K/\nu_K^2$, with $\zeta_K=\sqrt{M/r^3}$, $\nu_K=\sqrt{M/(r-2M)}$ and $\Gamma_K=\sqrt{r/(r-3M)}$.
The corresponding timelike conditions $|\nu_\pm|<1$ identify the allowed regions for the radial coordinate where co/counter-rotating geodesics exist, i.e., the location of the light-ring (LR)
\beq
r_{{(\rm LR)}\pm}
=2M\left\{1+\cos\left[\frac23\arccos\left(\mp\frac{a}{M}\right)\right]\right\}\,.
\eeq
The latter are the positive roots of the cubic equation $\Gamma_\pm^{-1}=0$, i.e.,
\beq
r^3-6Mr^2+9M^2r-4a^2M
=0\,.
\eeq

The dimensionless electric-type and magnetic-type tidal invariants in this case are given by
\begin{eqnarray}
\label{THgeo}
\tilde {\mathcal T}_E(U_{\pm})&=& 
\frac{6M^6}{r^6}+\tilde {\mathcal T}_B(U_{\pm})\,, \\
\tilde {\mathcal T}_B(U_{\pm})
&=&\frac{18M^6\Delta}{r^7}\left[\frac{\sqrt{Mr}\mp a}{\sqrt{r}(r-3M)\pm2a\sqrt{M}}\right]^2
\,.\nonumber
\end{eqnarray}
They both diverge as approaching the LR as $\sim1/(r-r_{{(\rm LR)}\pm})^2$. 
For instance, for $a/M=0.5$ one finds $\tilde {\mathcal T}_B(U_{+})\approx0.03 M^2 /(r-r_{{(\rm LR)}+})^2$ and $\tilde {\mathcal T}_B(U_{-})\approx0.02 M^2 /(r-r_{{(\rm LR)}-})^2$, with $r_{{(\rm LR)}+}\approx2.3473M$ and $r_{{(\rm LR)}-}\approx3.5321M$.

By introducing the Boyer-Lindquist inverse (dimensionless) radius
\beq
u=\frac{M}{r}\,,
\eeq
and the dimensionless rotation parameter ${\hat a}\equiv a/M$, Eq. (\ref{THgeo}) become
\begin{eqnarray}
\label{THgeou}
\tilde {\mathcal T}_E(U_{\pm})&=& \frac{6u^6}{(1-3u\mp2{\hat a}u^{3/2})^2}\left[1-3u+3u^2\right.\nonumber\\
&&\left.
\mp2{\hat a}u^{3/2}(1+{\hat a}^2u^2)+{\hat a}^2u^3(1+3{\hat a}^2u)\right]\,, \nonumber\\
\tilde {\mathcal T}_B(U_{\pm})&=&18u^7\frac{\tilde\Delta(1\mp{\hat a}\sqrt{u})^2}{(1-3u\mp2{\hat a}u^{3/2})^2}\,,
\end{eqnarray}
where 
\beq
\tilde\Delta=1-2u+{\hat a}^2u^2\,,
\eeq
and the factor in front of each fraction represents the Newtonian value of the tidal invariants, i.e., $\tilde {\mathcal T}_E^{\rm Newt}=6u^6$ and $\tilde {\mathcal T}_B^{\rm Newt}=18u^7$.

Let us consider the \lq\lq weak field" limit of the above expressions.
The expansion for $u\ll1$ gives
\begin{eqnarray}
\label{THgeoexp}
\tilde {\mathcal T}_E(U_{\pm})&=& 6u^6+\tilde {\mathcal T}_B(U_{\pm})\,, \nonumber\\
\tilde {\mathcal T}_B(U_{\pm})&=&\tilde{\mathcal T}_B^{(0)}+\tilde{\mathcal T}_B^{({\hat a})}{\hat a}+\tilde{\mathcal T}_B^{({\hat a}^2)}{\hat a}^2\,,
\end{eqnarray}
with
\begin{eqnarray}
\tilde{\mathcal T}_B^{(0)}&=&18u^7[1+4u+15u^2+54u^3+189u^4+O(u^5)]\,,\nonumber\\
\tilde{\mathcal T}_B^{({\hat a})}&=&\mp36u^{15/2}[1+6u+29u^2+126u^3+513u^4\nonumber\\
&&
+O(u^5)]\,,\nonumber\\
\tilde{\mathcal T}_B^{({\hat a}^2)}&=&18u^8[1+13u+89u^2+489u^3+O(u^4)]\,,
\end{eqnarray}
where terms higher than quadratic in the rotation parameter ${\hat a}$ have been neglected.

\subsection{General equatorial orbits}

Equatorial orbits have $4$-velocity given by
\beq
\label{orbitagen}
U
 =\gamma [n +\nu^{\hat r} e_{\hat r}+\nu^{\hat \phi} e_{\hat \phi}]
\ ,
\eeq
where the relative velocity $\nu(U,n)=\nu^{\hat r} e_{\hat r}+\nu^{\hat \phi} e_{\hat \phi}$ has magnitude $\nu=\sqrt{\nu_{\hat r}^2 +\nu_{\hat \phi}^2}$, with associated Lorentz factor $\gamma=(1-\nu^2)^{-1/2}$.
The parametric equations of the orbit are the solutions of the evolution equations $U=\rmd x^\alpha/\rmd\tau$, i.e., 
\begin{eqnarray}
\label{Ucompts}
\frac{\rmd t}{\rmd \tau}&=&\frac{\gamma}{N}\,,\qquad
\frac{\rmd r}{\rmd \tau} =\frac{\gamma\nu^{\hat r}}{\sqrt{g_{rr}}}\,,\nonumber\\
\frac{\rmd \phi}{\rmd \tau}&=& \frac{\gamma}{\sqrt{g_{\phi\phi}}}\left(\nu^{\hat \phi}-\frac{\sqrt{g_{\phi\phi}}N^\phi}{N}\right)\,.
\end{eqnarray}
Notice that for equatorial motion a convenient parametrization can be $r$ itself.
The relation (\ref{relTEH}) holds for an arbitrary equatorial orbit too, with $r=r(\tau)$ to be taken along the world line $U$, as already stated.

\section{Extended bodies with spin-induced quadrupolar deformations}

Hereafter, we will refer to the body 1 as an extended body (with mass $m_1=m$ and spin $S_1$) moving in a Kerr background (i.e., in the gravitational field of the body 2, with mass $m_2=M$ and spin $S_2=Ma$).
The dynamics of extended bodies in a given gravitational field is described by the MPD model \cite{math37,papa51,tulc59,dixon64,dixon69,dixon70,dixon73,dixon74}. 
In the quadrupole approximation, MPD equations read
\begin{eqnarray}
\label{papcoreqs}
\frac{{\rm D}P^{\mu}}{\rmd \tau} & = &
- \frac12 \, R^\mu{}_{\nu \alpha \beta} \, U^\nu \, S^{\alpha \beta}
-\frac16 \, \, J^{\alpha \beta \gamma \delta} \, \nabla^\mu R_{\alpha \beta \gamma \delta}
\,,\nonumber\\
\frac{{\rm D}S^{\mu\nu}}{\rmd \tau} & = & 
2 \, P^{[\mu}U^{\nu]}+
\frac43 \, J^{\alpha \beta \gamma [\mu}R^{\nu]}{}_{\gamma \alpha \beta}
\,,
\end{eqnarray}
where $P^{\mu}=m u^\mu$ (with $u \cdot u = -1$) is the total 4-momentum of the body with mass $m$, $S^{\mu \nu}$ is a (antisymmetric) spin tensor, $J^{\alpha\beta\gamma\delta}$ is the quadrupole tensor, and $U^\mu=\rmd z^\mu/\rmd\tau$ is the timelike unit tangent vector of the \lq\lq center of mass line'' (with parametric equations $x^\mu=z^\mu(\tau)$) used to make the multipole reduction, parametrized by the proper time $\tau$. 
Note that in general the mass $m$ is not constant along the world line of the extended body, and should be distinguished from the (constant) \lq\lq bare'' mass $m_0$.  
The tensor quantities introduced above are defined along the center of mass line only and all depend on $\tau$.

Additional constraints are imposed to the spin tensor \cite{tulc59,dixon64}
\beq
\label{tulczconds}
S^{\mu\nu}u{}_\nu=0\,,
\eeq
so that it is fully represented by a spatial vector (with respect to $u$), i.e.,
\beq
S(u)^\alpha=\frac12 \eta(u)^\alpha{}_{\beta\gamma}S^{\beta\gamma}=[{}^{*_{(u)}}S]^\alpha\,,
\eeq
where $\eta(u)_{\alpha\beta\gamma}=\eta_{\mu\alpha\beta\gamma}u^\mu$ is the spatial (with respect to $u$) unit volume 3-form with $\eta_{\alpha\beta\gamma\delta}=\sqrt{-g} \epsilon_{\alpha\beta\gamma\delta}$ the unit volume 4-form and $\epsilon_{\alpha\beta\gamma\delta}$ ($\epsilon_{0123}=1$) the Levi-Civita alternating symbol. As standard, hereafter we denote the spacetime dual of a tensor (built up with $\eta_{\alpha\beta\gamma\delta}$) by a $^*$, whereas the 
spatial dual of a spatial tensor with respect to $u$ (built up with $\eta(u)_{\alpha\beta\gamma}$) by $^{*_{(u)}}$.
It is also useful to introduce the signed magnitude $s$ of the spin vector
\beq
\label{sinv}
s^2=S(u)^\beta S(u)_\beta = \frac12 S_{\mu\nu}S^{\mu\nu}=-\frac12{\rm Tr}[S^2]\,, 
\eeq
which is in general not constant along the trajectory of the extended body. 

We will consider the special case of a quadrupole tensor completely determined by the spin structure of the body (see, e.g., Refs. \cite{steinhoff,steinhoff2}), i.e.,
\beq
\label{Jspininduced}
J^{\alpha\beta\gamma\delta}=4{u}^{[\alpha}\widetilde{\mathcal X}(u)^{\beta][\gamma}{u}^{\delta]}\,, \qquad
\widetilde{\mathcal X}(u)=\frac{C_Q}{m}[S^2]^{\rm STF}\,,
\eeq
where $C_Q$ is a constant parameter which is characteristic of the body under consideration (see, e.g., \cite{steinhoff2} with $C_Q\to (3/4)C_Q$) and $[S^2]^{\rm STF}$  
denotes the trace-free part of the square of the spin tensor, i.e., 
\begin{eqnarray}
[S^2]^{\rm STF}{}^{\alpha\beta}
&=&S(u)^{\alpha}S(u)^{\beta}-\frac13s^2P(u)^{\alpha\beta}\nonumber\\
&=&[S(u)\otimes S(u)]^{\rm STF}{}^{\alpha\beta}\,,
\end{eqnarray}
where both the spin vector and the associated spin invariant have been used and $P(u)$ projects orthogonally to $u$.

We refer to Refs. \cite{quadrupschw,quadrupkerr1,quadrupkerrnum,quadrupkerr2,steinhoff,steinhoff2,steinhoff3,hinderer} for recent applications to the Schwarzschild and Kerr spacetimes of the MPD model in the quadrupole approximation. 
Here, we use the results of Ref. \cite{quadrupkerr2}, where the trajectory of the extended body has been fully determined under the assumptions of equatorial plane motion and spin of the body aligned with the rotation axis of the Kerr source, i.e.,
\beq
S=S^{\hat\theta}e_{\hat \theta}=-se_{\hat \theta}
\,.
\eeq  
The MPD equations imply that in this case the spin magnitude remains constant during the evolution, while the mass $m$ varies with $r$ as 
\beq
m =  m_0\left[1-\frac23C_Q{\hat s}^2\Gamma_\pm^2y_\pm^3 (1\mp4{\hat a}u^{3/2}+3{\hat a}^2u^2)\right]\,,
\eeq
where the dimensionless spin parameter ${\hat s}=s/(m_0M)$ has been introduced.
Implicit in the MPD model is the assumption that the characteristic length scale associated with the spin structure of the body be small enough if compared with the characteristic length of the background curvature, i.e., $|{\hat s}|\ll1$, in order to avoid backreaction effects.
Note that along circular orbits the mass $m$ is also a constant.

\subsection{Circular orbits}

Let us consider first the special case in which the orbit of the extended body remains circular.
The 4-velocity $U=U_{\rm (circ)}$ is thus given by Eq. (\ref{orbita}) with normalization factor
\beq
\label{solGamma}
\Gamma=\Gamma_\pm\left(1+{\hat s}\Gamma_{\hat s}+{\hat s}^2\Gamma_{\hat s\hat s} \right)\,,
\eeq
with
\begin{eqnarray}
\Gamma_{\hat s}&=& 
-\frac32\Gamma_\pm^2y_\pm^3\nu_\pm\frac{\sqrt{\tilde\Delta}}{u}(1\mp {\hat a}\sqrt{u})\,,
\\
\Gamma_{\hat s\hat s}&=&
\mp\frac23C_Q\Gamma_{\hat s}\sqrt{u}\frac{\tilde\Delta\mp4{\hat a}u^{3/2}(1\mp {\hat a}\sqrt{u})}{1\mp {\hat a}\sqrt{u}}
 \nonumber\\
&&
\pm \frac38\Gamma_\pm^4y_\pm^{15/2}u^{-7/2}(1\mp {\hat a}\sqrt{u})
[10-21u+13{\hat a}^2u^2
\nonumber\\
&&
+5{\hat a}^2u^3+9{\hat a}^4u^4\pm4{\hat a}u^{3/2}(3-6u+7{\hat a}^2u^2)
]
 \,,\nonumber
\end{eqnarray}
and angular velocity  
\beq
\label{solzeta}
\zeta=\zeta_\pm\left(1+{\hat s}\zeta_{\hat s}+{\hat s}^2\zeta_{\hat s\hat s}  \right)\,,
\eeq
with
\begin{eqnarray}
\label{zetaquad}
\zeta_{\hat s}&=&
 -\frac32 y_\pm^{3/2}(1\mp {\hat a}\sqrt{u})\,,
\\
\zeta_{\hat s\hat s} &=& 
\mp\frac23C_Q\zeta_{\hat s}\sqrt{u}\frac{\tilde\Delta\mp4{\hat a}u^{3/2}(1\mp {\hat a}\sqrt{u})}{1\mp {\hat a}\sqrt{u}}\nonumber\\
&&
\pm\frac38y_\pm^{3}(1\mp {\hat a}\sqrt{u})
[7+9{\hat a}^2u^2
\pm {\hat a}\sqrt{u}(3+u)]
\,,\nonumber
\end{eqnarray}
respectively, where
\begin{eqnarray}
M\zeta_\pm&=&\frac{\pm u^{3/2}}{1\pm{\hat a}u^{3/2}}
\equiv y_\pm^{3/2}\,,\nonumber\\
\nu_\pm&=&\frac{y_\pm^{3/2}}{u\sqrt{\tilde\Delta}}(1+{\hat a}^2u^2\mp2{\hat a}u^{3/2})\,,\nonumber\\
\Gamma_\pm&=&\frac{1\pm{\hat a}u^{3/2}}{(1-3u\pm2{\hat a}u^{3/2})^{1/2}}\,.
\end{eqnarray}
The parametric equations of the orbit are then given by Eq. (\ref{casocircorb}).

The dimensionless electric-type and magnetic-type tidal invariants turn out to be 
\begin{eqnarray}
\label{THspincirc}
\tilde{\mathcal T}_E(U_{\rm(circ)})&=&\tilde{\mathcal T}_E(U_\pm)+\tilde{\mathcal T}_B(U_\pm)\left(\hat s \delta_{\hat s}+\hat s^2 \delta_{\hat s \hat s}\right)\,,
\nonumber\\
\tilde{\mathcal T}_B(U_{\rm(circ)})&=&\tilde{\mathcal T}_B(U_\pm)\left(1+\hat s \delta_{\hat s}+\hat s^2 \delta_{\hat s \hat s}
\right)\,,
\end{eqnarray}
so that
\begin{eqnarray}
\label{deltaT_EB}
\tilde{\mathcal T}_E(U_{\rm(circ)})-\tilde{\mathcal T}_E(U_\pm)&=&
\tilde{\mathcal T}_B(U_{\rm(circ)})-\tilde{\mathcal T}_B(U_\pm)\nonumber\\
&=&\tilde{\mathcal T}_B(U_\pm)\left(\hat s \delta_{\hat s}+\hat s^2 \delta_{\hat s \hat s}\right)\,,\qquad
\end{eqnarray}
with 
\begin{eqnarray}
\delta_{\hat s}
&=& \mp3u^{-3/2}\Gamma_\pm^2y_\pm^3[\tilde\Delta+u(1\mp {\hat a}\sqrt{u})^2]
\,,\nonumber\\
\delta_{\hat s \hat s}
&=&\mp\frac23C_Q\delta_{\hat s}\sqrt{u}\frac{\tilde\Delta\mp4{\hat a}u^{3/2}(1\mp {\hat a}\sqrt{u})}{1\mp {\hat a}\sqrt{u}}
\nonumber\\
&&
+\frac34\Gamma_\pm^4y_\pm^6u^{-3}
\{10-16u-2(3-22{\hat a}^2)u^2
\nonumber\\
&&
-40{\hat a}^2u^3+48{\hat a}^4u^4\pm{\hat a}\sqrt{u}[3-60u\nonumber\\
&&
+(97+6{\hat a}^2)u^2-86{\hat a}^2u^3]
\}\,,
\end{eqnarray}
and $\tilde{\mathcal T}_E(U_\pm)$ and $\tilde{\mathcal T}_B(U_\pm)$ given by Eq. (\ref{THgeou}).

In the weak field limit, the spin-induced corrections (\ref{deltaT_EB}) to the geodesic values $\tilde{\mathcal T}_{E,B}(U_\pm)$ (see Eq. (\ref{THgeoexp})) can be written as
\begin{eqnarray}
\tilde{\mathcal T}_E(U_{\rm(circ)})-\tilde{\mathcal T}_E(U_\pm)&=&
\tilde{\mathcal T}_B(U_{\rm(circ)})-\tilde{\mathcal T}_B(U_\pm)\\
&=&\tilde{\mathcal T}_B^{(\hat s)}{\hat s}+2\tilde{\mathcal T}_B^{({\hat a}{\hat s})}{\hat a}{\hat s}+\tilde{\mathcal T}_B^{(\hat s^2)}{\hat s}^2\,,\qquad
\nonumber
\end{eqnarray}
where
\begin{eqnarray}
\tilde{\mathcal T}_B^{(\hat s)}&=&-54u^{17/2}[1+6u+29u^2+O(u^3)]\,,\nonumber\\
\tilde{\mathcal T}_B^{({\hat a}{\hat s})}&=&54u^9[1+10u+O(u^2)]\,,\nonumber\\
\tilde{\mathcal T}_B^{({\hat s}^2)}&=&9u^9[4C_Q+u(15+16C_Q)+O(u^2)]\,.\qquad
\end{eqnarray}

\begin{figure}
\centering
\includegraphics[scale=0.35]{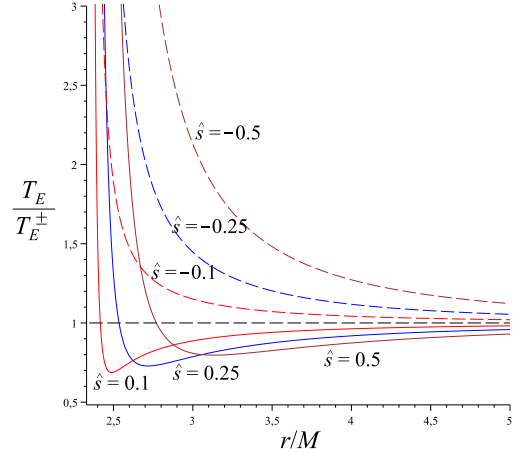}
\caption{
The behavior of the ratio ${\mathcal T}_E(U_{\rm(circ)})/{\mathcal T}_E(U_{\pm})$ is shown as a function of the radial coordinate for $a/M=0.5$, $C_Q=1$ and different values of the spin parameter ${\hat s}=[-0.5,-0.25,0,0.25,0.5]$ in the case of co-rotating reference circular geodesics.
The latter have their LR at $r_{{(\rm LR)}+}/M\approx2.347$, where it is positively divergent.
The values of $\hat s$ have been exaggerated to enhance the effect. 
}
\label{fig:1}
\end{figure}

We show in Fig.~\ref{fig:1} the behavior of the ratio ${\mathcal T}_E(U_{\rm(circ)})/{\mathcal T}_E(U_{\pm})$ as a function of the radial coordinate for selected values of the parameters. The allowed range of $r$ is determined by the existence of circular geodesics, i.e., $r>r_{{(\rm LR)}+}\approx2.347M$ in the co-rotating case.
The positive divergence at the LR is expected from geodesic motion. 
Spin corrections involve additional divergent terms in comparison with those already present in the $\Gamma$ factor (see Eq. (\ref{solGamma})); hence, in general, they enhance such a  behavior.
Our analysis shows that the value of the electric tidal invariant associated with the orbit of the extended body is always greater than the corresponding geodesic value if the spin vector is anti-aligned with the rotation axis of the background source.
In the aligned case, instead, it is exactly the opposite for large radii.
As the LR is approached, the curves exhibit a minimum, crossing then the geodesic value and finally indefinitely growing very close to the LR.


\subsubsection{Inverting the relation between $\zeta$ and $u$}

In order to get a gauge-invariant information from the tidal invariants computed above, one has to express them in terms of an observable quantity, like the angular velocity $\zeta$, instead of the radial coordinate.
We will use the following rescaled angular velocity 
\beq
\label{fdf_omega}
\zeta'=\frac{\zeta}{1-a\zeta}\,,
\eeq
first introduced in the literature in 1991 by de Felice and Usseglio-Tomasset in the context of a relativistically correct definition of strains and centrifugal forces in general relativity \cite{fdf1,fdf2}.
Later on it received much attention in Refs. \cite{idcf1,idcf2}, where its geometrical meaning in terms of new slicings of Kerr spacetime was clarified (see text after Eq. (4.4) of Ref. \cite{idcf2}).

Substituting Eq. (\ref{solzeta}) into Eq. (\ref{fdf_omega}) yields\footnote{
A prime over ${\zeta'}_{\hat s}$ and ${\zeta'}_{\hat s\hat s}$ should not be confused here with a derivative.
}
\beq
\label{zeta_primo2}
M\zeta'=\pm u^{3/2} (1+\hat s {\zeta'}_{\hat s}+{\hat s}^2 {\zeta'}_{\hat s\hat s})\,,
\eeq
where
\begin{eqnarray}
{\zeta'}_{\hat s}
&=& \mp \frac32 u^{3/2}(1\mp {\hat a}\sqrt{u})\,, \nonumber\\
{\zeta'}_{\hat s\hat s} 
&=& \frac38u^3(7\pm3{\hat a}\sqrt{u})(1\mp{\hat a}\sqrt{u})\nonumber\\
&& 
+C_Qu^2(1-2u+5{\hat a}^2u^2\mp4{\hat a}u^{3/2})\,,
\end{eqnarray}
with associated dimensionless variable 
\beq
y'\equiv (M\zeta')^{2/3}
=u\left[1+\frac23 {\hat s}{\zeta'}_{\hat s}+{\hat s}^2\left(\frac23 {\zeta'}_{\hat s\hat s} -\frac19 {\zeta'}_{\hat s}{}^2\right) \right]\,.
\eeq
Inverting then the above relation gives
\beq
\label{rel_uyp}
u=y'[1+\alpha(y'){\hat s}+\beta(y'){\hat s}^2]\,,
\eeq
with 
\begin{eqnarray}
\alpha(y')
&=&   \pm y'{}^{3/2}(1\mp{\hat a}y'{}^{1/2})\,, \nonumber\\
\beta(y')
&=&   -\frac13  y'{}^2[-3y'(1\mp5{\hat a}\sqrt{y'}+4{\hat a}^2y')\nonumber\\
&& 
+2C_Q(1-2y'+5{\hat a}^2y'{}^2\mp4{\hat a}y'{}^{3/2})] \,. \qquad
\end{eqnarray}
Finally, the relation between $u$ and $\zeta$ can be in turn easily obtained from the previous equation by replacing $y'$ in terms of $\zeta$ through Eq. (\ref{fdf_omega}).

Alternatively, one can directly use the variable $y$ introduced in Eq. (\ref{ydef}) and related to $y'$ by 
\begin{eqnarray}
\label{ypvsy}
y'&=&\frac{y}{(1-\hat a y^{3/2})^{2/3}}\nonumber\\
&=& y\left[1+\frac23{\hat a}y^{3/2}+\frac59{\hat a}^2y^3+O(y^{9/2})\right]\,.
\end{eqnarray}
The relation with $u$ in this case is 
\begin{eqnarray}
\label{rel_uy}
u&=&\frac{y}{(1-\hat a y^{3/2})^{2/3}}(1+\tilde \alpha (y){\hat s}+\tilde \beta (y){\hat s}^2)\\
&=&y\left\{1+\left(\frac23{\hat a}\pm{\hat s}\right)y^{3/2}-{\hat s}\left({\hat a}+\frac23C_Q{\hat s}\right)y^2\right.\nonumber\\
&&\left.
+\left[\frac59{\hat a}^2\pm\frac53{\hat a}{\hat s}+\left(1+\frac43C_Q\right){\hat s}^2\right]y^3+O(y^{7/2})\right\}\,,
\nonumber
\end{eqnarray}
where
\begin{eqnarray}
\tilde \alpha (y)&=&\alpha\left(\frac{y}{(1-\hat a y^{3/2})^{2/3}}\right)\,,\nonumber\\
\tilde \beta (y)&=&\beta\left(\frac{y}{(1-\hat a y^{3/2})^{2/3}}\right)\,.
\end{eqnarray}

\subsubsection{Gauge-invariant expressions of tidal invariants}

Let us turn to the general expressions (\ref{THspincirc}) of tidal invariants.
In terms of the gauge-invariant quantity $y'$ introduced above they read 
\begin{eqnarray}
\tilde{\mathcal T}_E(U_{\rm(circ)})\big|_{y'}&=&\tilde{\mathcal T}_E(U_\pm)\big|_{u=y'}(1+{\hat s}w^{E}_{\hat s}+{\hat s}^2w^{E}_{\hat s\hat s})\,,\nonumber\\
\tilde{\mathcal T}_B(U_{\rm(circ)})\big|_{y'}&=&\tilde{\mathcal T}_B(U_\pm)\big|_{u=y'}(1+{\hat s}w^{B}_{\hat s}+{\hat s}^2w^{B}_{\hat s\hat s})\,,\qquad
\nonumber\\
\end{eqnarray}
where
\begin{eqnarray}%
w^{E}_{\hat s}&=&\pm6(1\mp{\hat a}\sqrt{y'})y'{}^{3/2}\frac{n_E}{d_E}\,,\nonumber\\
w^{E}_{\hat s\hat s}&=&
\mp\frac23C_Qw^{E}_{\hat s}\sqrt{y'}\frac{1-2y'+5{\hat a}^2y'{}^2\mp4{\hat a}y'{}^{3/2}}{1\mp{\hat a}\sqrt{y'}}\nonumber\\
&&
+\frac{3y'{}^3(1\mp{\hat a}\sqrt{y'})}{1-3y'\pm2{\hat a}y'{}^{3/2}}\frac{b_E}{d_E}\,,\nonumber\\
w^{B}_{\hat s}&=&\frac{\pm2y'{}^{3/2}n_B}{1-2y'+{\hat a}^2y'{}^2}\,,\nonumber\\
w^{B}_{\hat s\hat s}&=&\mp\frac23C_Qw^{B}_{\hat s}\sqrt{y'}\frac{1-2y'+5{\hat a}^2y'{}^2\mp4{\hat a}y'{}^{3/2}}{1\mp{\hat a}\sqrt{y'}}\nonumber\\
&&
+\frac{y'{}^3b_B}{(1-2y'+{\hat a}^2y'{}^2)(1-3y'\pm2{\hat a}y'^{3/2})}\,,\qquad
\end{eqnarray}
with
\begin{eqnarray}%
n_E&=&
d_E-y'(1-{\hat a}^2y'{}^2)(1-y'+2{\hat a}^2y'{}^2\mp2{\hat a}y'{}^{3/2})
\,,\nonumber\\
d_E&=&1-3y'+3(1+{\hat a}^2)y'{}^2+{\hat a}^2y'{}^3+3{\hat a}^4y'{}^4\nonumber\\
&&
\mp2{\hat a}y'{}^{3/2}(1+3{\hat a}^2y'{}^2)\,,\nonumber\\
b_E&=&7-53y'+9(15+4{\hat a}^2)y'{}^2-3(37+55{\hat a}^2)y'{}^3\nonumber\\
&&
+5{\hat a}^2(31+11{\hat a}^2)y'{}^4-545{\hat a}^4y'{}^5-170{\hat a}^6y'{}^6\nonumber\\
&&
\mp{\hat a}\sqrt{y'}
[13-95y'{}+3(79+20{\hat a}^2)y'{}^2\nonumber\\
&&
-9(21+31{\hat a}^2)y'{}^3+{\hat a}^2(111+85{\hat a}^2)y'{}^4\nonumber\\
&&
-599{\hat a}^4y'{}^5]
\,,
\end{eqnarray}
and
\begin{eqnarray}%
n_B&=&
2-5y'+3{\hat a}^2y'{}^2\mp{\hat a}^2\sqrt{y'}(4-9y'+5{\hat a}^2y'{}^2)
\,,\nonumber\\
b_B&=&10-6(9-10{\hat a}^2)y'+3(26-135{\hat a}^2)y'{}^2\nonumber\\
&&
+(647+85{\hat a}^2){\hat a}^2y'{}^3-429{\hat a}^4y'{}^4\nonumber\\
&&
\mp2{\hat a}\sqrt{y'}[27-156y'+3(77-5{\hat a}^2)y'{}^{2}-6{\hat a}^2y'{}^3\nonumber\\
&&
-85{\hat a}^4y'{}^4]\,.
\end{eqnarray}
The corresponding (weak field) $y'-$expansions are
\begin{eqnarray}%
\tilde{\mathcal T}_E(U_{\rm(circ)})\big|_{y'}&=&6y'{}^6[1+3y'\mp6({\hat a}-{\hat s})y'{}^{3/2}\nonumber\\
&&
+(12+3{\hat a}^2-6{\hat a}{\hat s}-4C_Q{\hat s}^2)y'{}^2\nonumber\\
&&
\mp12(3{\hat a}-{\hat s})y'{}^{5/2}+O(y'{}^3)]\,,\nonumber\\
\tilde{\mathcal T}_B(U_{\rm(circ)})\big|_{y'}&=&18y'{}^7\left[1\mp2{\hat a}\sqrt{y'}+(4+{\hat a}^2)y'\right.\nonumber\\
&&\left.
\mp4(3{\hat a}-{\hat s})y'{}^{3/2}
+\left(15+13{\hat a}^2-16{\hat a}{\hat s}\right.\right.\nonumber\\
&&\left.\left.
-\frac83C_Q{\hat s}^2\right)y'{}^2
+O(y'{}^{5/2})\right]\,.
\end{eqnarray}
Noticeably, the dependence of the above result on the rotational parameter $\hat a$ is exact (indeed the expansion has involved the $y'$ variable only). 

In the Schwarzschild case ($y'=y$) the previous expressions for the electric and magnetic tidal invariants simplify to

\begin{widetext}

\begin{eqnarray}
\tilde{\mathcal T}_E(U_{\rm(circ)})\big|_{y}&=&
6y^6\left\{\frac{1-3y+3y^2}{(1-3y)^2}
\pm6y^{3/2}\frac{1-4y+4y^2}{(1-3y)^2}{\hat s}\right.\nonumber\\
&&\left.
-\frac{3y^2}{(1-3y)^2}\left[
\frac43C_Q(1-2y)^3-\frac{y}{1-3y}(7-53y+135y^2-111y^3)
\right]{\hat s}^2
\right\}\nonumber\\
&=&
6y^6[1+3y\pm6{\hat s}y^{3/2}+4(3-C_Q{\hat s}^2)y^2
\pm12{\hat s}y^{5/2}+O(y^3)]\,,\nonumber\\
\tilde{\mathcal T}_B(U_{\rm(circ)})\big|_{y}&=&
18y^7\left\{\frac{1-2y}{(1-3y)^2}
\pm2y^{3/2}\frac{2-5y}{(1-3y)^2}{\hat s}\right.\nonumber\\
&&\left.
-\frac{2y^2}{(1-3y)^2}\left[
\frac23C_Q(1-2y)(2-5y)-\frac{y}{1-3y}(5-27y+39y^2)
\right]{\hat s}^2
\right\}\nonumber\\
&=&
18y^7\left[1+4y
\pm4{\hat s}y^{3/2}
+\left(15-\frac83C_Q{\hat s}^2\right)y^2
+O(y^{5/2})\right]
\,.
\end{eqnarray}

\end{widetext}
In the absence of spin, these results agree with the test-mass limit ($q\to0$) of Eq. (\ref{eq:2.8}), where one sets ${\mathcal M}=M(1+q)\to M$, $X_1\to0$, $X_2\to1$ and $x\to y$. 
The spin-dependent terms are genuinely new and can be used to compare with similar results obtained through numerical approaches.

\begin{figure}
\centering
\includegraphics[scale=0.35]{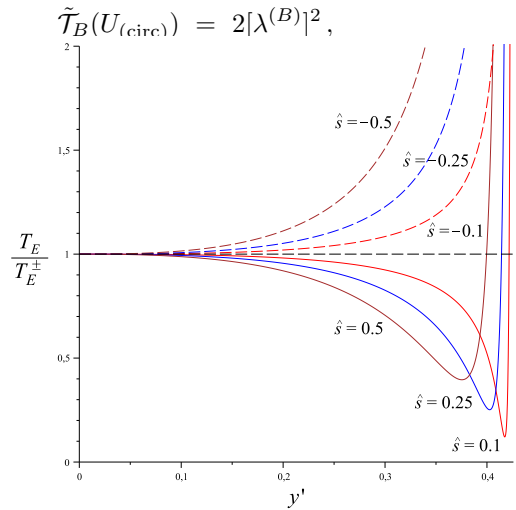}
\caption{
The behavior of the ratio ${\mathcal T}_E(U_{\rm(circ)})/{\mathcal T}_E(U_{\pm})$ is shown as a function of the dimensionless rescaled angular velocity variable $y'$ for the same choice of parameters as in Fig.~\ref{fig:1}. 
The allowed range for $y'$ is between $0$ and $My'_+\approx0.426$, associated with $r>r_{{(\rm LR)}+}$.
}
\label{fig:2}
\end{figure}

In Fig.~\ref{fig:2} we show the behavior of the ratio ${\mathcal T}_E(U_{\rm(circ)})/{\mathcal T}_E(U_{\pm})$ as a function of the dimensionless rescaled angular velocity $y'$, for selected values of the parameters. 
The advantage of using the angular velocity as the independent variable in place of the radial coordinate is that the former is a gauge-invariant quantity.  Fig.~\ref{fig:2} exactly reproduces the features discussed in Fig.~\ref{fig:1} in both cases of aligned/anti-aligned spin, which are then general enough.


\subsubsection{Tidal eigenvalues}

Let us evaluate the eigenvalues of the electric and magnetic tidal tensors $M^2{\mathcal E}^{\mu}{}_{\nu}(U_{\rm(circ)})$ and $M^2{\mathcal B}^{\mu}{}_{\nu}(U_{\rm(circ)})$.
They are such that \cite{Bini:2014zxa,Dolan:2014pja}
\begin{eqnarray} 
{\mathcal E}(U_{\rm(circ)})&=&{\rm diag}[\lambda_1^{(E)},\lambda_2^{(E)},-(\lambda_1^{(E)}+\lambda_2^{(E)})]\,,\nonumber\\
{\mathcal B}(U_{\rm(circ)})&=&{\rm diag}[\lambda^{(B)},-\lambda^{(B)},0]\,,
\end{eqnarray} 
where we have taken into account their tracelessness and the existence of a zero magnetic eigenvalue (see also Appendix \ref{eigenvec}, where the corresponding eigenvectors are also given).
The three independent eigenvalues can be computed by using the quadratic invariants $\tilde{\mathcal T}_E(U_{\rm(circ)})$ and $\tilde{\mathcal T}_B(U_{\rm(circ)})$ and the cubic invariant ${\rm Tr}\,[{\mathcal E}(U_{\rm(circ)})]^3$, which are related by 
\begin{eqnarray} 
\tilde{\mathcal T}_B(U_{\rm(circ)})&=&2[\lambda^{(B)}]{}^2\,,\nonumber\\
\frac12\tilde{\mathcal T}_E(U_{\rm(circ)})&=&[\lambda_1^{(E)}]{}^2+[\lambda_2^{(E)}]{}^2+\lambda_1^{(E)}\lambda_2^{(E)}\,,\nonumber\\
\frac13M^6{\rm Tr}\,[{\mathcal E}(U_{\rm(circ)})]^3&=&-\lambda_1^{(E)}\lambda_2^{(E)}(\lambda_1^{(E)}+\lambda_2^{(E)})\,.
\end{eqnarray} 
The cubic invariant turns out to be given by

\begin{widetext}

\begin{eqnarray}
M^6{\rm Tr}\,[{\mathcal E}(U_{\rm(circ)})]^3&=&-3u^3\Gamma_\pm^4y_\pm^6(1+3{\hat a}^2u^2\mp4{\hat a}u^{3/2})(2-3u+3{\hat a}^2u^2\mp2{\hat a}u^{3/2})\nonumber\\
&&
\pm81u^{5/2}\Gamma_\pm^6y_\pm^9\tilde\Delta(1\mp{\hat a}\sqrt{u})^2(1-u+2{\hat a}^2u^2\mp2{\hat a}u^{3/2}){\hat s}\nonumber\\
&&
-\frac{27}{4}u\Gamma_\pm^8y_\pm^{12}\tilde\Delta(1\mp{\hat a}\sqrt{u})\left\{
8C_Q\frac{u^2}{\Gamma_\pm^2y_\pm^3}(1-u+2{\hat a}^2u^2
\mp2{\hat a}u^{3/2})(1-2u+5{\hat a}^2u^2\mp4{\hat a}u^{3/2})\right.\nonumber\\
&&\left.
+3(1\mp{\hat a}\sqrt{u})[10-16u+2u^2(22{\hat a}^2-3)-40{\hat a}^2u^3+48{\hat a}^4u^4\right.\nonumber\\
&&\left.
\pm{\hat a}\sqrt{u}(3-60u+u^2(6{\hat a}^2+97)-86{\hat a}^2u^3)]
\right\}{\hat s}^2\,.
\end{eqnarray} 

\end{widetext}
Noticeably, the first and second order spin corrections of the electric cubic invariant are simply related to the corresponding ones of the electric quadratic invariant by 
\beq
\frac{\frac13M^6{\rm Tr}\,[{\mathcal E}(U_{\rm(circ)})]^3{}^{(1)}}{\frac12\tilde{\mathcal T}_E(U_{\rm(circ)}){}^{(1)}}
=-u^3
=\frac{\frac13M^6{\rm Tr}\,[{\mathcal E}(U_{\rm(circ)})]^3{}^{(2)}}{\frac12\tilde{\mathcal T}_E(U_{\rm(circ)}){}^{(2)}}\,.
\eeq

\begin{figure}
\centering
\includegraphics[scale=0.35]{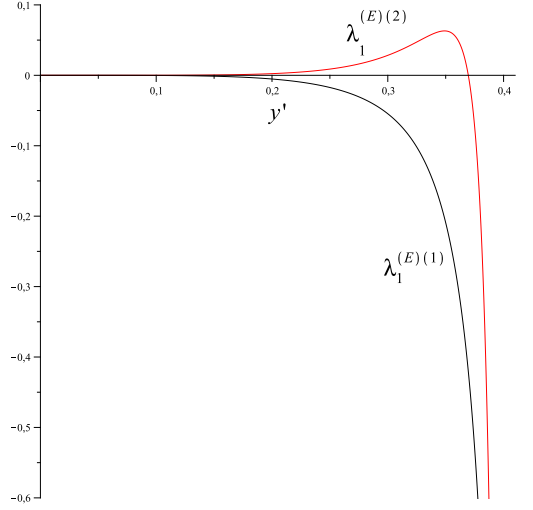}
\caption{
The behavior of both first and second order corrections in spin of the electric eigenvalue $\lambda_1^{(E)}$ is shown as a function of the rescaled angular velocity variable $y'$ for $a/M=0.5$ and $C_Q=1$ in the case of co-rotating reference circular geodesics.
}
\label{fig:3}
\end{figure}

The above eigenvalues have the following form
\begin{eqnarray} 
\label{eigendef}
\lambda^{(B)}&=&\lambda^{(B)\,(0)}+{\hat s}\lambda^{(B)\,(1)}+{\hat s}^2\lambda^{(B)\,(2)}\,,\nonumber\\
\lambda_{1,2}^{(E)}&=&\lambda_{1,2}^{(E)\,(0)}+{\hat s}\lambda_{1,2}^{(E)\,(1)}+{\hat s}^2\lambda_{1,2}^{(E)\,(2)}\,,
\end{eqnarray}
with
\begin{eqnarray} 
-\lambda^{(B)\,(0)}&=&
-3u^{7/2}\sqrt{\tilde\Delta}\frac{1\mp{\hat a}\sqrt{u}}{1-3u\pm2{\hat a}u^{3/2}}\,,\nonumber\\
\lambda^{(B)\,(1)}&=&
\frac12\lambda^{(B)\,(0)}\delta_{\hat s}\,,\nonumber\\
\lambda^{(B)\,(2)}&=&
\frac12\lambda^{(B)\,(0)}\left(\delta_{\hat s \hat s}-\frac{\delta_{\hat s}^2}{4}\right)\,,
\end{eqnarray}
and
\begin{eqnarray}%
\lambda_1^{(E)\,(0)}&=&
u^{3}-\frac{3u^3\tilde\Delta}{1-3u\pm2{\hat a}u^{3/2}}\,,\nonumber\\
\lambda_2^{(E)\,(0)}&=&
-2u^{3}+\frac{3u^3\tilde\Delta}{1-3u\pm2{\hat a}u^{3/2}}\,,\nonumber\\
\lambda_1^{(E)\,(1)}&=&-\lambda_2^{(E)\,(1)}
=\pm9u^{11/2}\frac{\tilde\Delta(1\mp{\hat a}\sqrt{u})^2}{(1-3u\pm2{\hat a}u^{3/2})^2}\,,\nonumber\\
\lambda_1^{(E)\,(2)}&=&-\lambda_2^{(E)\,(2)}\nonumber\\
&=&-6u^6\frac{\tilde\Delta(1\mp{\hat a}\sqrt{u})}{(1-3u\pm2{\hat a}u^{3/2})^2}
\left\{
C_Q(1-2u\right.\nonumber\\
&&\left.
+5{\hat a}^2u^2\mp4{\hat a}u^{3/2})
+\frac38\frac{u(1\mp{\hat a}\sqrt{u})}{1-3u\pm2{\hat a}u^{3/2}}
[10\right.\nonumber\\
&&\left.
-18u+18{\hat a}^2u^2\pm{\hat a}\sqrt{u}(3-13u)]
\right\}\,.
\end{eqnarray}
Therefore, the value of the third electric eigenvalue $\lambda_3^{(E)}=-(\lambda_1^{(E)}+\lambda_2^{(E)})$ is not modified to second order in spin with respect to the corresponding geodesic value $\lambda_3^{(E)\,(0)}=u^3$.
The behavior of both first and second order corrections in spin of the electric eigenvalue $\lambda_1^{(E)}$ as a function of $y'$ is shown in Fig.~\ref{fig:3}, as an example.
They are both mostly negligible for small values of $y'$ (weak field), but their contribution becomes more and more significant as approaching the LR.


To conclude this section on circular motion, it is interesting to compare the spin-induced corrections to the tidal eigenvalues to first order in spin in the limiting case of a Schwarzschild spacetime with the self-force corrections obtained in Ref. \cite{Bini:2014zxa}, i.e.,
\begin{eqnarray}
\lambda^{(B)}&=&\lambda^{(B)\,(0)}+q\lambda^{(B)\,1SF}\,,\nonumber\\
\lambda_{1,2}^{(E)}&=&\lambda_{1,2}^{(E)\,(0)}+q\lambda_{1,2}^{(E)\,1SF}\,,
\end{eqnarray}
with 
\begin{eqnarray}
-\lambda^{(B)\,1SF}&=&
-2y^{7/2}\left[
1+\frac32y+\frac{59}{8}y^2+O(y^3)
\right]\,,\nonumber\\
\lambda_1^{(E)\,1SF}&=&
2y^3\left[
1+y-\frac{19}{4}y^2+O(y^3)
\right]
\,,\nonumber\\
\lambda_2^{(E)\,1SF}&=&
-y^3\left[
1+\frac32y+\frac{23}{8}y^2+O(y^3)
\right]
\,.
\end{eqnarray}
The solutions (\ref{eigendef}) for the tidal eigenvalues in terms of the coordinate variable $u$ evaluated at $\hat a=0$ become
\begin{eqnarray} 
-\lambda^{(B)}&=&
-3u^{7/2}\frac{\sqrt{1-2u}}{1-3u}\left\{1
\mp\frac32u^{3/2}\frac{1-u}{1-3u}{\hat s}\right.\nonumber\\
&&\left.
+\frac18\frac{u^{2}}{(1-3u)^2}
[8C_Q(1-u)(1-2u)(1-3u)\right.\nonumber\\
&&\left.
+3u(7-10u-9u^2)]{\hat s}^2
\right\}\,,\nonumber\\
\lambda_1^{(E)}&=&
-u^{3}\frac{2-3u}{1-3u}
\pm9u^{11/2}\frac{1-2u}{(1-3u)^2}{\hat s}\nonumber\\
&&
-\frac32u^6\frac{1-2u}{(1-3u)^3}
[4C_Q(1-2u)(1-3u)\nonumber\\
&&+3u(5-9u)]{\hat s}^2
\,,\nonumber\\
\lambda_2^{(E)}&=&
\frac{u^3}{1-3u}
\mp9u^{11/2}\frac{1-2u}{(1-3u)^2}{\hat s}\nonumber\\
&&
+\frac32u^6\frac{1-2u}{(1-3u)^3}
[4C_Q(1-2u)(1-3u)\nonumber\\
&&
+3u(5-9u)]{\hat s}^2
\,,
\end{eqnarray}
or, in terms of the gauge-invariant variable $y$ defined in Eq. (\ref{ydef}),

\begin{widetext}

\begin{eqnarray}
-\lambda^{(B)}&=&
-3y^{7/2}\frac{\sqrt{1-2y}}{1-3y}\left\{1
\pm y^{3/2}\frac{2-5y}{1-2y}{\hat s}
+y^2\left[
-\frac23C_Q(2-5y)
+y\frac{6-42y+101y^2-81y^3}{2(1-2y)^2(1-3y)}
\right]{\hat s}^2
\right\}\,,\nonumber\\
\lambda_1^{(E)}&=&
-y^3\frac{2-3y}{1-3y}\left\{1
\pm 3y^{3/2}\frac{2-5y}{2-3y}{\hat s}
+\frac{y^2}{2-3y}\left[
-2C_Q(2-5y)(1-2y)
+\frac{3y}{1-3y}(4-23y+36y^2)
\right]{\hat s}^2
\right\}\,,\nonumber\\
\lambda_2^{(E)}&=&
\frac{y^3}{1-3y}\left\{1
\pm3y^{3/2}(1-2y){\hat s}
+y^2\left[
-2C_Q(1-2y)^2
+\frac{3y}{1-3y}(2-11y+18y^2)
\right]{\hat s}^2
\right\}\,,
\end{eqnarray}
whose weak field expansion gives
\begin{eqnarray}
-\lambda^{(B)}&=&
-3y^{7/2}\left\{
1+2y\pm2{\hat s}y^{3/2}+\left(\frac{11}{2}-\frac43C_Q{\hat s}^2\right)y^2
\pm3{\hat s}y^{5/2}
+\left[16+\left(3+\frac23C_Q\right){\hat s}^2\right]y^3
+O(y^{7/2})
\right\}\,,\nonumber\\
\lambda_1^{(E)}&=&
-2y^3\left\{
1+\frac32y\pm3{\hat s}y^{3/2}+\left(\frac{9}{2}-2C_Q{\hat s}^2\right)y^2
\pm\frac32{\hat s}y^{5/2}
+\left[\frac{27}{2}+3(2+C_Q){\hat s}^2\right]y^3
+O(y^{7/2})
\right\}
\,,\nonumber\\
\lambda_2^{(E)}&=&
y^3\left\{
1+3y\pm3{\hat s}y^{3/2}+\left(9-2C_Q{\hat s}^2\right)y^2
\pm3{\hat s}y^{5/2}
+\left[27+2(3+C_Q){\hat s}^2\right]y^3
+O(y^{7/2})
\right\}
\,.
\end{eqnarray}

\end{widetext}
Therefore, the ratio between 1SF corrections and spin-induced corrections behaves as $y^{3/2}$ for all eigenvalues.
In fact, to first order in spin (1S) the latter are given by
\begin{eqnarray}
-\lambda^{(B)\,1S}&=&
\mp6y^5\left[1+\frac32y+\frac72y^2+O(y^3)\right]
\,,\nonumber\\ 
\lambda_1^{(E)\,1S}&=&
\mp6y^{9/2}\left[1+\frac12y+\frac32y^2+O(y^3)\right]
\,,\nonumber\\
\lambda_2^{(E)\,1S}&=&
\pm3y^{9/2}\left[1+y+3y^2+O(y^3)\right]
\,.
\end{eqnarray}

\subsection{Quasi-circular orbits}

Let us consider the solution to the MPD equations corresponding to a quasi-circular orbit with unit tangent vector $U^\mu=\rmd x^\mu/\rmd\tau$, i.e., the initial conditions being chosen so that the world line of the extended body has the same starting point as the reference circular geodesic at radius $r=r_0$ for vanishing spin.
We also require that the two world lines are initially tangent. 

The complete solution (up to $O(\hat s^2)$ included) is given by \cite{quadrupkerr1,quadrupkerr2}
\beq
x^\alpha=x^\alpha_\pm+{\hat s}x^\alpha_{(1)}+{\hat s}^2x^\alpha_{(2)}\,,
\eeq
with
\begin{eqnarray} 
\label{solord1}
\tilde t_{(1)}&\equiv&\Omega_{\rm(ep)}t_{(1)}
=\tilde T_{\hat s}(\sin\ell-\ell)\,,\nonumber\\
\tilde r_{(1)}&\equiv&\frac{r_{(1)}}{r_0}
=\tilde R_{\hat s}(\cos\ell-1)\,,
\nonumber\\
\tilde \phi_{(1)}&\equiv&(M\Omega_{\rm(ep)})\phi_{(1)}
=(M\bar\zeta_\pm)\tilde t_{(1)}\,,
\end{eqnarray}
and
\begin{eqnarray} 
\label{solord2}
\tilde t_{(2)}&\equiv&\Omega_{\rm(ep)}t_{(2)}\nonumber\\
&=&\tilde D_1\sin\ell+\tilde D_2\sin2\ell+D_3\ell\cos\ell+D_4\ell\,, \nonumber\\
\tilde r_{(2)}&\equiv&\frac{r_{(2)}}{r_0}\nonumber\\
&=&\tilde C_1(\cos\ell-1)+\tilde C_2(\cos2\ell-1)+\tilde C_3\ell\sin\ell\,, \nonumber\\
\phi_{(2)}&=&E_1\sin\ell+E_2\sin2\ell+\tilde E_3\ell\cos\ell+\tilde E_4\ell\,.
\end{eqnarray}
Here $\ell\equiv\Omega_{\rm(ep)}\tau$ is a parameter along the orbit and
\beq
\label{omegaep}
M\Omega_{\rm(ep)}=
\Gamma_\pm y_\pm^{3/2}\left[1-6u_0-3{\hat a}^2u_0^2\pm 8{\hat a}u_0^{3/2}\right]^{1/2}
\eeq
denotes the well known epicyclic frequency governing the radial perturbations of circular geodesics, with associated dimensionless variable defined as
\beq
y_{\rm(ep)}\equiv\left(\frac{M\Omega_{\rm(ep)}}{\Gamma_\pm}\right)^{2/3}
\,.
\eeq
The explicit expressions for the coefficients are listed in Appendix \ref{const}.

The unit tangent vector to the orbit is then given by
\beq
U^\alpha=U^\alpha_{\pm}+{\hat s}U^\alpha_{(1)}+{\hat s}^2U^\alpha_{(2)}\,,
\eeq
with
\beq
U_{(1)}=\gamma_\pm{\mathcal V}^{(r)}_{\hat s}\left[2\frac{u_0^{3/2}}{\Gamma_\pm y_{\rm(ep)}^{3/2}}(\cos\ell-1)\bar U_\pm
+\sin\ell e_{\hat r}\right]\,,
\eeq
and
\beq
U_{(2)}=XU_\pm+Y\bar U_\pm+Ze_{\hat r}\,,
\eeq
with
\begin{eqnarray}
X&=&-\frac12\gamma_\pm^2({\mathcal V}^{(r)}_{\hat s})^2[\cos2\ell-2\cos\ell+1]
\,,\\
Y&=&{\bar\Gamma_\pm}^{-1}[-\Gamma_\pm X+
2\tilde D_2\cos2\ell+(D_3+\tilde D_1)\cos\ell\nonumber\\
&&
-D_3\ell\sin\ell+D_4]
\,,\nonumber\\
Z&=&
\frac{\Gamma_\pm y_{\rm(ep)}^{3/2}}{u_0\sqrt{\tilde\Delta}}[-2\tilde C_2\sin2\ell+(\tilde C_3-\tilde C_1)\sin\ell+\tilde C_3\ell\cos\ell]
\,.\nonumber
\end{eqnarray}

\subsubsection{Tidal invariants along quasi-circular orbits} 

\begin{figure*}
\centering
\subfigure[]{\includegraphics[scale=0.35]{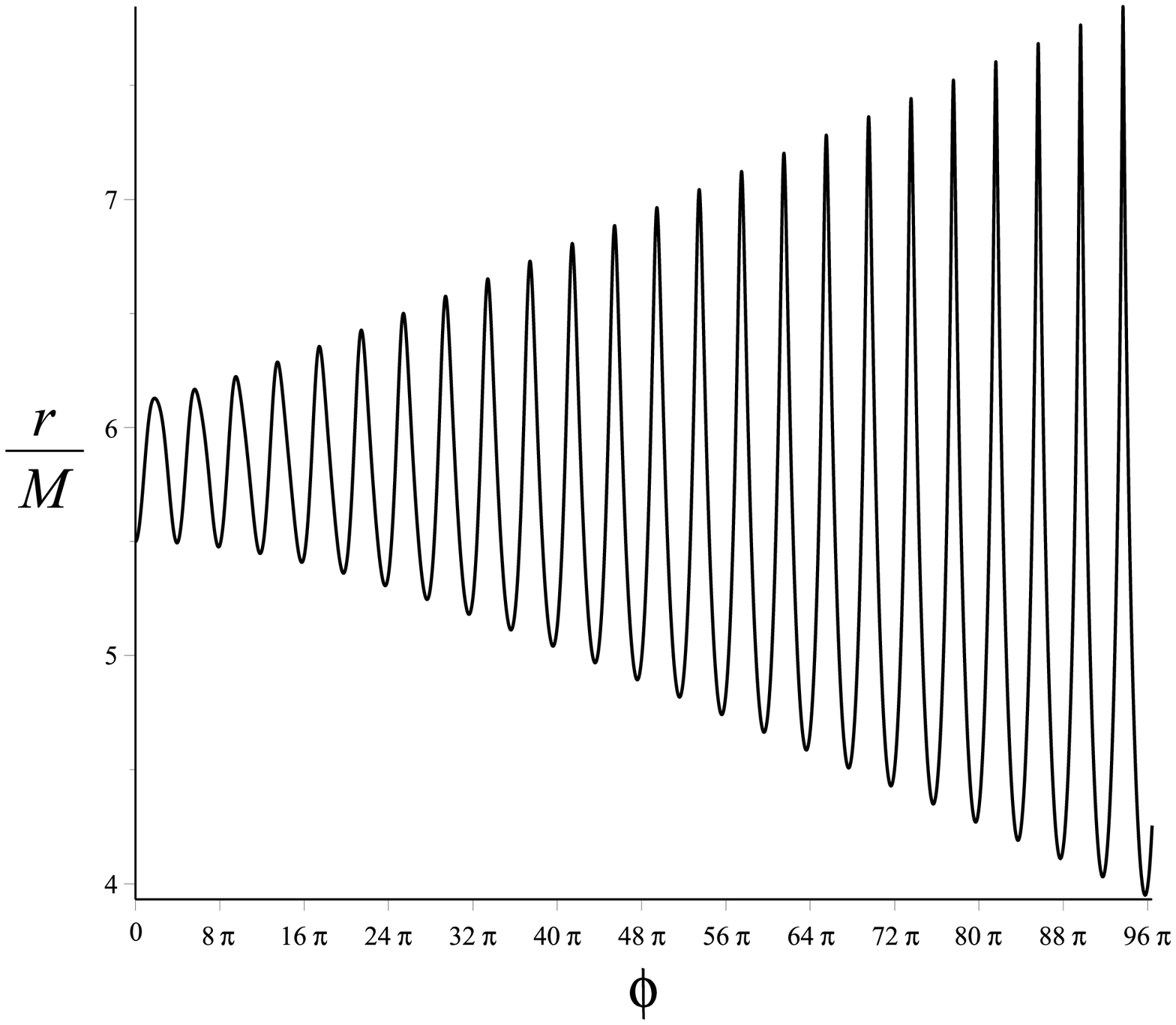}}
\hspace{5mm}
\subfigure[]{\includegraphics[scale=0.35]{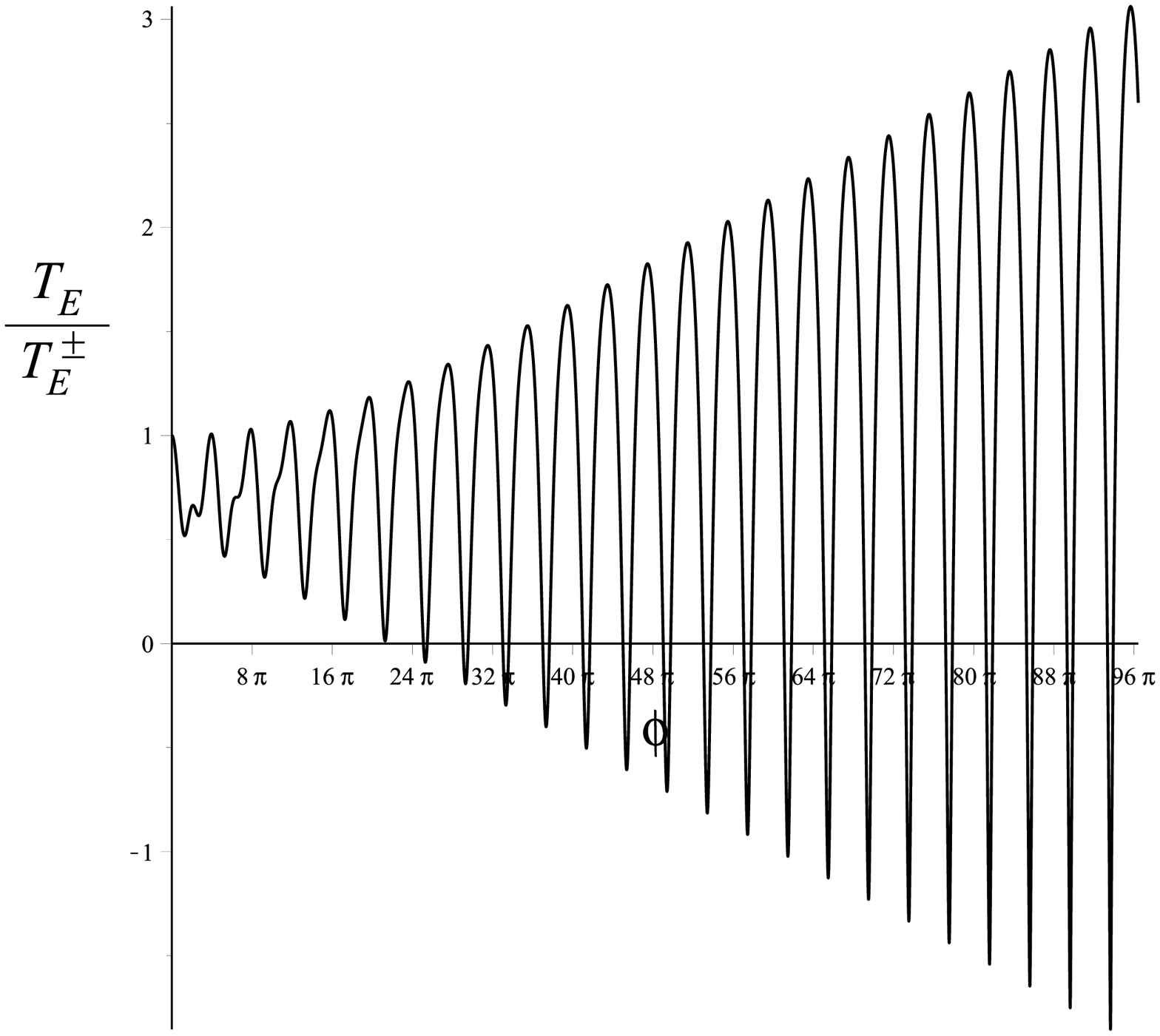}}
\caption{
An example of non-circular motion is shown in the panel (a), where the evolution of the radial coordinate $r/M$ is plotted as a function of $\phi$ for $a/M=0.5$, ${\hat s}=0.1$ and $C_Q=1$.
The reference orbit is chosen as the co-rotating circular geodesic at $r_0/M=5.5$. 
The corresponding behavior of the ratio ${\mathcal T}_E(U)/{\mathcal T}_E(U_{\pm})$ is shown in the panel (b).
Note that the value of the spin parameter has been exaggerated to enhance the effect.
It should actually be very small in order that the MPD model be valid, implying that the orbit would remain \lq\lq quasi circular'' after a large number of revolutions.
}
\label{fig:4}
\end{figure*}

The tidal electric and magnetic invariants can be written as
\beq
\tilde{\mathcal T}_{E,B}(U)=\tilde{\mathcal T}_{E,B}(U_\pm)+{\hat s}\tilde{\mathcal T}^{(1)}_{E,B}+{\hat s}^2\tilde{\mathcal T}^{(2)}_{E,B}
\,.
\eeq
It turns out that the first and second order corrections to the tidal electric and magnetic invariants are related by
\begin{eqnarray}
\tilde{\mathcal T}^{(1)}_E-\tilde{\mathcal T}^{(1)}_B&=&-36u_0^6\tilde r_{(1)}
\,,\nonumber\\
\tilde{\mathcal T}^{(2)}_E-\tilde{\mathcal T}^{(2)}_B&=&-36u_0^6\left(\tilde r_{(2)}-\frac72\tilde r_{(1)}^2\right)
\,.
\end{eqnarray}
We find
\begin{eqnarray}
\label{THspingen}
\tilde{\mathcal T}_E(U)&=&\tilde{\mathcal T}_E(U_\pm)
+\left(\hat s \epsilon^E_{\hat s}+\hat s^2 \epsilon^E_{\hat s \hat s}\right)(\cos\ell-1)\nonumber\\
&&
+\hat s^2 \zeta^E_{\hat s \hat s}(\cos2\ell-1)
+\hat s^2 \eta^E_{\hat s \hat s}\ell\sin\ell\,,
\nonumber\\
\tilde{\mathcal T}_B(U)&=&\tilde{\mathcal T}_B(U_\pm)
+\left(\hat s \epsilon^B_{\hat s}+\hat s^2 \epsilon^B_{\hat s \hat s}\right)(\cos\ell-1)\nonumber\\
&&
+\hat s^2 \zeta^B_{\hat s \hat s}(\cos2\ell-1)
+\hat s^2 \eta^B_{\hat s \hat s}\ell\sin\ell\,,
\end{eqnarray}
with 
\begin{eqnarray}%
\epsilon^E_{\hat s}&=&\epsilon^B_{\hat s}-36u_0^6\tilde R_{\hat s}\,,\nonumber\\
\epsilon^E_{\hat s \hat s}&=&\epsilon^B_{\hat s \hat s}-36u_0^6(\tilde C_1+7\tilde R_{\hat s}^2)\,,\nonumber\\
\zeta^E_{\hat s \hat s}&=&\zeta^B_{\hat s \hat s}-36u_0^6\left(\tilde C_2-\frac74\tilde R_{\hat s}^2\right)\,,\nonumber\\
\eta^E_{\hat s \hat s}&=&\eta^B_{\hat s \hat s}-36u_0^6\tilde C_3\,.
\end{eqnarray}
The first order correction is 
\begin{eqnarray}
\label{epsilonBdef}
\epsilon^B_{\hat s}&=&
\pm108\Gamma_\pm^4\frac{y_\pm^9}{y_{\rm(ep)}^3}u_0^{5/2}(1\mp {\hat a}\sqrt{u_0})^3\tilde\Delta[4\tilde\Delta\nonumber\\
&&
+u_0(1\mp {\hat a}\sqrt{u_0})^2]
\,,
\end{eqnarray}
whereas the second order corrections have very long expressions and are listed in Appendix \ref{const}.
We list below their approximate expansions in the weak field limit:
\begin{eqnarray}%
\epsilon^B_{\hat s \hat s}&=&
\pm432u_0^{17/2}\left[
1\mp3{\hat a}\sqrt{u_0}+\left(\frac{33}{4}+3{\hat a}^2\right)u_0\right.\nonumber\\
&&\left.
+O(u_0^{3/2})\right]
\,,\nonumber\\
\epsilon^B_{\hat s \hat s}&=&
-288u_0^9\left\{
C_Q(1\mp2{\hat a}\sqrt{u_0})\right.\nonumber\\
&&\left.
+\left[\frac{75}{2}+C_Q\left(\frac{25}{4}+{\hat a}^2\right)\right]u_0
+O(u_0^{3/2})\right\}
\,,\nonumber\\
\zeta^B_{\hat s \hat s}&=&
3249u_0^{10}\left[
1\mp\frac{39}{10}{\hat a}\sqrt{u_0}+\left(\frac{563}{40}+\frac{57}{10}{\hat a}^2\right)u_0\right.\nonumber\\
&&\left.
+O(u_0^{3/2})\right]
\,,\nonumber\\
\eta^B_{\hat s \hat s}&=&
=1296u_0^{10}\left[
1\mp4{\hat a}\sqrt{u_0}+\left(\frac{37}{4}+6{\hat a}^2\right)u_0\right.\nonumber\\
&&\left.
+O(u_0^{3/2})\right]
\,.
\end{eqnarray}
Fig.~\ref{fig:4} shows the evolution of the radial coordinate as well as the behavior of the ratio ${\mathcal T}_E(U)/{\mathcal T}_E(U_{\pm})$ along the orbit for selected values of the parameters. 
The orbit of the extended body initially deviates slightly from the reference circular geodesic, but after a number of oscillations it gets closer and closer to the central source. 
This is due to the chosen value of the spin parameter, which has been exaggerated to enhance the effect.
The secular term appearing at the second order in spin is responsible for driving the growth of the oscillations after each revolution.
Note that the amplitude of the oscillations should actually maintain small in order that the orbit be \lq\lq quasi-circular.'' 
This is achieved by choosing a smaller value of ${\hat s}$, which should be indeed much less than 1 to preserve the validity of the MPD model, as already stated.

A final remark concerns the behavior of tidal eigenvalues.
Their expressions in this case are quite long and not very illuminating, so that we avoid showing them.
However, the structure of the solution is similar to that of tidal invariants evaluated above, containing oscillating terms with frequency equal to the epicyclic frequency and a secular term enhancing the amplitude of the oscillations.


\section{Concluding remarks}

We have computed the lowest order tidal invariants (of both electric and magnetic types) along the (non-geodesic) world line of an extended body in the equatorial plane of a Kerr spacetime.
The body is spinning and tidally deformed. Its motion is described in terms of a number of (scalar, vector, tensor) fields defined along a single world line (\lq\lq center of mass'' line). These fields are associated with a multipolar expansion around the center of mass line according to the MPD model truncated at the quadrupole order. The quadrupole tensor is assumed to be quadratic in spin, accounting for deformations induced by the spin itself. We have discussed the behavior of such invariants when the body is moving along a circular orbit as well as in the case of arbitrary (equatorial) motion. The resulting expressions are in general quite involved, so we have also computed the corresponding weak field expansion, which can be directly compared with the existing PN literature, especially in the limiting case of the Schwarzschild solution.

In the circular case the results are expressed in a gauge-invariant way in terms of a suitably defined dimensionless angular velocity. The spin dependent terms are genuinely new and can provide useful information when compared with purely numerical investigations. 
We have shown that first order spin corrections to tidal invariants appear at the $1.5$PN fractional accuracy beyond the leading (Newtonian) order, as expected. Quadratic spin corrections, instead, arise at the $3$PN fractional accuracy.
The behavior at the light ring is dominated by the corresponding geodesic values, i.e., the tidal invariants of both kinds diverge as $(r-r_{\rm(LR)})^{-2}$.
We have also evaluated the eigenvalues of the electric and magnetic tidal tensors, whose derivation requires the computation of the cubic electric invariant too. In the simpler Schwarzschild situation we have compared the self force expressions in absence of spin (0S-1SF) of the tidal eigenvalues with their counterparts in the case in which one takes into account spin corrections but neglects those due to self force (1S-0SF).
We have found that the ratio between 0S-1SF corrections and 1S-0SF behaves as $y^{3/2}$ for eigenvalues of both kind. This simple example may give some useful information concerning the interplay of the associated effects.

In the non-circular case, instead, the electric and magnetic invariants (as well as the associated tidal eigenvalues) vary along the orbit, exhibiting an oscillating behavior with frequency equal to the epicyclic frequency and twice its value plus a secular term which appears at the second order in spin only.
A possible extension of the present analysis would be that of converting the Boyer-Lindquist coordinate tidal information into some more efficient formalism when dealing with a two-body system, like the EOB formalism.  
We leave this further analysis to future works.

\appendix

\section{Circular orbits: eigenvectors of tidal tensors}
\label{eigenvec}

We compute in this section the eigenvectors associated with the electric and magnetic tidal tensors in the case of circular motion (see also Ref. \cite{Dolan:2014pja}).

Let us first introduce the orthonormal frame adapted to $U_{\rm (circ)}$ (see also Section \ref{circ})
\begin{eqnarray}
\label{eleframe}
E_0&\equiv& U_{\rm (circ)}=\gamma [n + \nu e_{\hat \phi}]\,, \quad
E_1=e_{\hat r}\,, \quad
E_2=e_{\hat \theta}\,, \nonumber\\
E_3&\equiv& \bar U_{\rm (circ)}={\rm sgn}(\nu)\gamma [\nu n +e_{\hat \phi}]\,, 
\end{eqnarray}
where $\gamma=(1-\nu^2)^{-1/2}$.
For example, if $\nu>0$, then $\{E_{a}\}$ is such that $\eta(U_{\rm (circ)})_{\alpha\beta\gamma}E_1^{\alpha}E_2^{\beta}E_3^{\gamma}=1$.

When decomposed with respect to that frame the electric tidal tensor is already diagonal with components 
\beq
{\mathcal E}(U_{\rm(circ)})={\rm diag}[\lambda_1^{(E)},\lambda_2^{(E)},\lambda_3^{(E)}]\,,
\eeq
so that $E_1$, $E_2$ and $E_3$ are the electric tidal eigenvectors associated with the eigenvalues
\begin{eqnarray} 
\lambda_1^{(E)}&=&\gamma^2[{\mathcal E}(n)_{\hat r\hat r}-2\nu{\mathcal B}(n)_{\hat r\hat \theta}-\nu^2{\mathcal E}(n)_{\hat \theta\hat \theta}]
\,,\nonumber\\
\lambda_2^{(E)}&=&\gamma^2[{\mathcal E}(n)_{\hat \theta\hat \theta}+2\nu{\mathcal B}(n)_{\hat r\hat \theta}-\nu^2{\mathcal E}(n)_{\hat r\hat r}]
\,,\nonumber\\
\lambda_3^{(E)}&=&-(\lambda_1^{(E)}+\lambda_2^{(E)})
=-({\mathcal E}(n)_{\hat r\hat r}+{\mathcal E}(n)_{\hat \theta\hat \theta})\,.
\end{eqnarray} 
Here ${\mathcal E}(n)$ and ${\mathcal B}(n)$ are the electric and magnetic parts of the Riemann tensor with respect to ZAMOs, whose nonvanishing components are given by
\begin{eqnarray}
\label{E_H}
{\mathcal E}(n)_{\hat r \hat r}&=& -\frac{M(2 r^4+5 r^2 a^2-2 a^2 M r+3 a^4)}{r^4 (r^3+a^2 r+2 a^2 M)}\,, \nonumber\\
{\mathcal E}(n)_{\hat \theta \hat \theta}&=&  \frac{M (r^4+4 r^2 a^2-4 a^2 M r+3 a^4)}{r^4 (r^3+a^2 r+2 a^2 M)}\nonumber\\
&=&-\frac{M}{r^3}- {\mathcal E}(n)_{\hat r \hat r}\,,\nonumber\\
{\mathcal B}(n)_{\hat r \hat \theta}&=&  -\frac{3 M a (r^2+a^2) \sqrt{\Delta}}{r^4 (r^3+a^2 r+2 a^2 M)}\,,
\end{eqnarray}
and ${\mathcal E}(n)_{\hat \phi \hat \phi}=-{\mathcal E}(n)_{\hat r \hat r}-{\mathcal E}(n)_{\hat \theta \hat \theta}=M/r^3$.
Therefore, the third eigenvalue $\lambda_3^{(E)}=M/r^3$ is not affected by the presence of the spin.
The solution for the linear velocity $\nu$ can be easily evaluated from that for $\zeta$ given by Eqs. (\ref{solzeta})--(\ref{zetaquad}), being simply related by Eq. (\ref{zetavsnu}), i.e.,  
\beq
\nu=\frac{\sqrt{g_{\phi\phi}}}{N} (\zeta+N^{\phi})
=\nu_\pm+\frac{\sqrt{g_{\phi\phi}}}{N}\zeta_\pm{\hat s}\left(\zeta_{\hat s}+{\hat s}\zeta_{\hat s\hat s}  \right)
\,.
\eeq

The magnetic tidal tensor, instead, is not diagonal with respect to the frame (\ref{eleframe}). However, it is enough to rotate the frame vectors $E_1$ and $E_2$ in the $r$-$\theta$ 2-plane to obtain 
\beq
{\mathcal B}(U_{\rm(circ)})={\rm diag}[\lambda^{(B)},-\lambda^{(B)},0]\,,
\eeq
with eigenvectors $(E_1\mp E_2)/\sqrt{2}$ and $E_3$ and associated eigenvalues $\pm\lambda^{(B)}$ and $0$, respectively, and
\beq
\lambda^{(B)}=\gamma^2[({\mathcal E}(n)_{\hat r\hat r}-{\mathcal E}(n)_{\hat \theta\hat \theta})\nu-(1+\nu^2){\mathcal B}(n)_{\hat r\hat \theta}]\,.
\eeq

\section{Quasi-circular orbits: explicit solution}
\label{const}

\subsection{Representation of the orbit}

We list below the various coefficients entering the quasi-circular orbit solution (\ref{solord1}) and (\ref{solord2}):
\begin{eqnarray} 
\tilde R_{\hat s}&\equiv&\frac{R_{\hat s}}{r_0} 
= -N\frac{u_0\sqrt{\tilde\Delta}}{y_{\rm(ep)}^{3/2}}{\mathcal V}^{(r)}_{\hat s}
\,,\nonumber\\
\tilde T_{\hat s}&\equiv&	\Omega_{\rm(ep)}T_{\hat s}
=\pm2\nu_\pm\frac{u_0^{3/2}}{y_{\rm(ep)}^{3/2}}{\mathcal V}^{(r)}_{\hat s}\,,
\end{eqnarray} 
with
\begin{eqnarray} 
N&=&\frac{\sqrt{\tilde\Delta}}{(1+{\hat a}^2u_0^2+2{\hat a}^2u_0^3)^{1/2}}\,,\nonumber\\
{\mathcal V}^{(r)}_{\hat s}&=&\pm3\frac{\sqrt{u_0\tilde\Delta}}{N}\frac{y_\pm^3}{y_{\rm(ep)}^{3/2}}(1\mp {\hat a}\sqrt{u_0})\,,
\end{eqnarray} 
and

\begin{widetext}

\begin{eqnarray} 
\label{solord2coeffs}
\tilde C_1&\equiv&\frac{C_1}{r_0}
=Nu_0\sqrt{\tilde\Delta}\frac{\nu_\pm}{y_{\rm(ep)}^{3/2}}
(B_1-\tilde B_3)
-9N^2u_0^3\frac{y_\pm^6}{y_{\rm(ep)}^6}(1\mp {\hat a}\sqrt{u_0})^2(1+{\hat a}^2u_0^2)[\tilde\Delta-u_0(1-{\hat a}^2u_0^2)]
\,,\nonumber\\
\tilde C_2&\equiv&\frac{C_2}{r_0}
=\frac12Nu_0\sqrt{\tilde\Delta}\frac{\nu_\pm}{y_{\rm(ep)}^{3/2}}
B_2
+\frac94N^2u_0^3\frac{y_\pm^6}{y_{\rm(ep)}^6}(1\mp {\hat a}\sqrt{u_0})^2(1+{\hat a}^2u_0^2)[\tilde\Delta-u_0(1-{\hat a}^2u_0^2)]
\,,\nonumber\\
\tilde C_3&\equiv&\frac{C_3}{r_0\Omega_{\rm(ep)}}
=-Nu_0\sqrt{\tilde\Delta}\frac{\nu_\pm}{y_{\rm(ep)}^{3/2}}\tilde B_3
\,,\nonumber\\
\tilde D_{1}&\equiv&\Omega_{\rm(ep)}D_1
=\mp2\gamma_\pm\nu_\pm^2\frac{u_0^{3/2}}{y_{\rm(ep)}^{3/2}}(B_{1}-2\tilde B_{3})
\mp9u_0^{5/2}\frac{N}{\tilde\Delta}\gamma_\pm\frac{y_\pm^{15/2}}{y_{\rm(ep)}^6}(1\mp {\hat a}\sqrt{u_0})^2
[7-12u_0+17{\hat a}^2u_0^2-28{\hat a}^2u_0^3\nonumber\\
&&
-(12-13{\hat a}^2){\hat a}^2u_0^4-12{\hat a}^4u_0^5-(44-3{\hat a}^2){\hat a}^4u_0^6+4{\hat a}^6u_0^7
\mp4{\hat a}u_0^{3/2}(4-9u_0+4{\hat a}^2u_0^2-2{\hat a}^2u_0^3-12{\hat a}^2u_0^4-{\hat a}^4u_0^5)]
\,,\nonumber\\
\tilde D_{2}&\equiv&\Omega_{\rm(ep)}D_2
=\mp\frac12\gamma_\pm\nu_\pm^2\frac{u_0^{3/2}}{y_{\rm(ep)}^{3/2}}B_{2}
\mp\frac94u_0^{5/2}\frac{N}{\tilde\Delta}\gamma_\pm\frac{y_\pm^{15/2}}{y_{\rm(ep)}^6}(1\mp {\hat a}\sqrt{u_0})^2
\{3-3u_0+9{\hat a}^2u_0^2-9{\hat a}^2u_0^3\nonumber\\
&&
+9{\hat a}^4u_0^4-{\hat a}^4u_0^5-(16-3{\hat a}^2){\hat a}^4u_0^6+5{\hat a}^6u_0^7
\mp2{\hat a}u_0^{3/2}[6-9u_0+8{\hat a}^2u_0^2+2{\hat a}^2u_0^3-2(6-{\hat a}^2){\hat a}^2u_0^4+3{\hat a}^4u_0^5]
\}\,,\nonumber\\
D_{3}&=&
\mp2\gamma_\pm\nu_\pm^2\frac{u_0^{3/2}}{y_{\rm(ep)}^{3/2}}\tilde B_{3}
\,,\qquad
D_{4}=
-\tilde D_{1}-2\tilde D_{2}-D_{3}
\,,\nonumber\\
E_{1}&=&
\frac{\bar y_\pm^{3/2}}{\Gamma_\pm y_{\rm(ep)}^{3/2}} \tilde D_{1}
-\frac{N^2u_0}{\nu_\pm\sqrt{\tilde\Delta}y_{\rm(ep)}^{3/2}}({\mathcal V}^{(r)}_{\hat s})^2
\,,\qquad
E_{2}=
\frac{\bar y_\pm^{3/2}}{\Gamma_\pm y_{\rm(ep)}^{3/2}} \tilde D_{2}
+\frac14\frac{N^2u_0}{\nu_\pm\sqrt{\tilde\Delta}y_{\rm(ep)}^{3/2}}({\mathcal V}^{(r)}_{\hat s})^2\,,\nonumber\\
\tilde E_3&\equiv&\frac{E_3}{\Omega_{\rm(ep)}}
=\frac{\bar y_\pm^{3/2}}{\Gamma_\pm y_{\rm(ep)}^{3/2}} \tilde D_{3}
\,,\qquad
\tilde E_4\equiv\frac{E_4}{\Omega_{\rm(ep)}}
=\frac{\bar y_\pm^{3/2}}{\Gamma_\pm y_{\rm(ep)}^{3/2}} D_{4}
+\frac12\frac{N^2u_0}{\nu_\pm\sqrt{\tilde\Delta}y_{\rm(ep)}^{3/2}}({\mathcal V}^{(r)}_{\hat s})^2\,,
\end{eqnarray}
with 
\beq
\bar y_\pm^{3/2}\equiv M\bar\zeta_\pm
=\pm\sqrt{u_0}\frac{1-2u_0\pm{\hat a}u_0^{3/2}}{1\pm2{\hat a}u_0^{3/2}+{\hat a}^2u_0^2}\,,
\eeq
and
\begin{eqnarray} 
B_{1}&=&
2C_Q\frac{u_0\sqrt{\tilde\Delta}}{N\nu_\pm}\frac{y_\pm^3}{y_{\rm(ep)}^{3/2}}[\tilde\Delta\mp4{\hat a}u_0^{3/2}(1\mp {\hat a}\sqrt{u_0})]
+3\frac{u_0^2N}{\sqrt{\tilde\Delta}\nu_\pm}\frac{y_\pm^9}{y_{\rm(ep)}^{15/2}}(1\mp {\hat a}\sqrt{u_0})
\{1-6u_0+4(3+4{\hat a}^2)u_0^2-69{\hat a}^2u_0^3\nonumber\\
&&
-3(12+17{\hat a}^2){\hat a}^2u_0^4+4(33-17{\hat a}^2){\hat a}^2u_0^5+22(2-3{\hat a}^2){\hat a}^4u_0^6-165{\hat a}^6u_0^7\nonumber\\
&&
\mp {\hat a}u_0^{3/2}[2+9{\hat a}^2u_0-103{\hat a}^2u_0^2-2(28+3{\hat a}^2){\hat a}^2u_0^3
+12(21-8{\hat a}^2){\hat a}^2u_0^4-3(68+5{\hat a}^2){\hat a}^4u_0^5-39{\hat a}^6u_0^6]
\}\,,\nonumber\\
B_{2}&=&
-\frac92\frac{u_0^2N}{\sqrt{\tilde\Delta}\nu_\pm}\frac{y_\pm^9}{y_{\rm(ep)}^{15/2}}(1\mp {\hat a}\sqrt{u_0})^2
[1+7u_0-(22+3{\hat a}^2)u_0^2+37{\hat a}^2u_0^3-(44+9{\hat a}^2){\hat a}^2u_0^4-(8-21{\hat a}^2){\hat a}^2u_0^5\nonumber\\
&&
+(34-5{\hat a}^2){\hat a}^4u_0^6-9{\hat a}^6u_0^7
\mp {\hat a}u_0^{3/2}(1+2u_0+{\hat a}^2u_0^2)(1-3u_0+{\hat a}^2u_0^2+{\hat a}^2u_0^3)]
\,,\nonumber\\
\tilde B_3&\equiv&\frac{B_3}{\Omega_{\rm(ep)}}
=9\frac{u_0^2\sqrt{\tilde\Delta}}{N\nu_\pm}\frac{y_\pm^9}{y_{\rm(ep)}^{15/2}}(1\mp {\hat a}\sqrt{u_0})^2
[1+11u_0+(18-7{\hat a}^2)u_0^2+31{\hat a}^2u_0^3
\pm8{\hat a}u_0^{3/2}(2-5u_0-{\hat a}^2u_0^2)]
\,.
\end{eqnarray}

\end{widetext}

\subsection{Representation of tidal invariants}

We list below the second order corrections to the tidal invariants (\ref{THspingen}):
\begin{eqnarray}
\epsilon^B_{\hat s \hat s}&=&
\mp\frac23C_Q\epsilon^B_{\hat s}\sqrt{u_0}\frac{\tilde\Delta\mp4{\hat a}u_0^{3/2}(1\mp {\hat a}\sqrt{u_0})}{1\mp {\hat a}\sqrt{u_0}}\nonumber\\
&&
\mp108u_0^4\Gamma_\pm^4N^2\frac{y_\pm^{15}}{y_{\rm(ep)}^9}(1\mp {\hat a}\sqrt{u_0})^3P_1(u_0)
\,,\nonumber\\
\zeta^B_{\hat s \hat s}&=&
\pm81u_0^4\Gamma_\pm^4N^2\frac{y_\pm^{15}}{y_{\rm(ep)}^9}(1\mp {\hat a}\sqrt{u_0})^3P_2(u_0)
\,,\nonumber\\
\eta^B_{\hat s \hat s}&=&
\pm\frac13\epsilon^B_{\hat s}\frac{N\nu_\pm}{\sqrt{u_0\tilde\Delta}}\frac{y_{\rm(ep)}^{3/2}}{y_\pm^{3}}\frac{\tilde B_3}{1\mp a\sqrt{u_0}}
\,,
\end{eqnarray}
where $\epsilon^B_{\hat s}$ is given by Eq. (\ref{epsilonBdef}) and

\begin{widetext}

\begin{eqnarray}%
P_1(u_0)&=&
100-892u_0+6(355-47{\hat a}^2)u_0^2-6(242+427{\hat a}^2)u_0^3+2{\hat a}^2(2613-1559{\hat a}^2)u_0^4+2{\hat a}^2(1800-641{\hat a}^2)u_0^5\nonumber\\
&&
-6{\hat a}^2(664-477{\hat a}^2+877{\hat a}^4)u_0^6+2{\hat a}^4(3902-2207{\hat a}^2)u_0^7+2{\hat a}^6(2651-1263{\hat a}^2)u_0^8-5346{\hat a}^8u_0^9\nonumber\\
&&
\mp{\hat a}\sqrt{u_0}
[84-1277u_0+(2794+147{\hat a}^2)u_0^2-(5941{\hat a}^2+1032)u_0^3-{\hat a}^2(-7508+435{\hat a}^2)u_0^4\nonumber\\
&&
-{\hat a}^2(-6884+8231{\hat a}^2)u_0^5-{\hat a}^2(1023{\hat a}^4+2210{\hat a}^2+5448)u_0^6-{\hat a}^4(-16948+4611{\hat a}^2)u_0^7\nonumber\\
&&
-3{\hat a}^6(2196+175{\hat a}^2)u_0^8-1140{\hat a}^8u_0^9]
\,,\nonumber\\
P_2(u_0)&=&
40-317u_0+(692-187{\hat a}^2)u_0^2-(428+241{\hat a}^2)u_0^3-{\hat a}^2(1191{\hat a}^2-700)u_0^4+{\hat a}^2(1508+617{\hat a}^2)u_0^5\nonumber\\
&&
-{\hat a}^2(-808{\hat a}^2+1661{\hat a}^4+1216)u_0^6-7{\hat a}^4(129{\hat a}^2-200)u_0^7-{\hat a}^6(697{\hat a}^2-2520)u_0^8-1444{\hat a}^8u_0^9\nonumber\\
&&
\mp{\hat a}\sqrt{u_0}
[36-503u_0+(1172+11{\hat a}^2)u_0^2-(1659{\hat a}^2+680)u_0^3-3(-720{\hat a}^2+87{\hat a}^4-56)u_0^4\nonumber\\
&&
-{\hat a}^2(-2332+1849{\hat a}^2)u_0^5-{\hat a}^2(411{\hat a}^4+2072+544{\hat a}^2)u_0^6-81{\hat a}^4(-60+13{\hat a}^2)u_0^7\nonumber\\
&&
-{\hat a}^6(1172+175{\hat a}^2)u_0^8-360{\hat a}^8u_0^9]
\,.
\end{eqnarray}

\end{widetext}

\section*{Acknowledgements}
DB thanks Prof. T. Damour for useful suggestions. 
The authors are indebted to Prof. G. Faye for fruitful discussions on spin-induced quadrupole interaction.
ICRANet and the Italian INFN (Section of Naples) are also acknowledged for partial support.

\end{document}